\documentclass[preprint,12pt]{elsarticle}
\usepackage[margin=1 in]{geometry}

\usepackage{booktabs}
\usepackage{graphicx}
\usepackage{upgreek}
\usepackage{multicol,multirow}
\usepackage{amsmath,amssymb,amsfonts}
\usepackage{mathrsfs}
\usepackage{amsthm}
\usepackage[figuresright]{rotating}
\usepackage{appendix}
\usepackage{ifpdf}
\usepackage[T1]{fontenc}
\usepackage{newtxtext}
\usepackage{newtxmath}
\usepackage{textcomp}
\usepackage{xcolor}
\usepackage{natbib}
\usepackage[normalem]{ulem}

\usepackage{amsmath,amsfonts,amssymb,mathtools}
\usepackage{gensymb}
\usepackage{siunitx}
\usepackage{cancel}

\journal{Journal of Fluids and Structures}

\begin{document}

\begin{frontmatter}

\title{Experimental identification of blade-level forces, torque, and pitching moment for cross-flow turbines} 

\author[label1]{Abigale Snortland}
\author[label2]{Katherine Van Ness}
\author[label3]{Jennifer A. Franck}
\author[label4]{Ari Athair}
\author[label4]{Owen Williams}
\author[label5]{Brian Polagye}

\affiliation[label1]{organization={Pacific Northwest National Lab},
            addressline={1100 Dexter Ave N},
            city={Seattle},
            postcode={98109},
            state={WA},
            country={USA}}

\affiliation[label2]{organization={University of Washington Applied Physics Lab},
            addressline={1013 NE 40th St},
            city={Seattle},
            postcode={98105},
            state={WA},
            country={USA}}

\affiliation[label3]{organization={University of Wisconsin-Madison Department of Mechanical Engineering},
            addressline={1500 Engineering Dr},
            city={Madison},
            postcode={53706},
            state={WI},
            country={USA}}

\affiliation[label4]{organization={University of Washington Department of Aeronautics and Astronautics},
            addressline={3940 Benton Lane NE},
            city={Seattle},
            postcode={98105},
            state={WA},
            country={USA}}

\affiliation[label5]{organization={University of Washington Department of Mechanical Engineering},
            addressline={3900 E Stevens Way NE},
            city={Seattle},
            postcode={98195},
            state={WA},
            country={USA}}

\textbf{This article may be downloaded for personal use only. This article appeared in (Abigale Snortland, Katherine Van Ness, Jennifer A. Franck, Ari Athair, Owen Williams, Brian Polagye,
Experimental identification of blade-level forces, torque, and pitching moment for cross-flow turbines,
Journal of Fluids and Structures,
Volume 138, 2025, ISSN 0889-9746) and may be found at (https://doi.org/10.1016/j.jfluidstructs.2025.104403).}

\begin{abstract}
Cross-flow turbine power is a net sum of power generation from rotating blades and power loss from rotating support structures. While the aggregate forces and torques at the turbine level are important for end use, these can inhibit a deeper understanding of fluid-structure interactions. Identification of blade-level forces and torques allows for specific investigations into how the fluid forcing on the blade drives rotation and can aid blade structural design. Here, we present a physics-based methodology for extracting blade-level forces and torques from experimental measurements at the axis of rotation of a cross-flow turbine, and demonstrate strong agreement with equivalent blade-only simulations. In doing so, we highlight the often-overlooked pitching moment, which offsets continuous increases in power generation from the tangential force and leads to net-zero power generation at freewheel.
\end{abstract}


Research highlights
\begin{highlights}
\item  A physics-based method is developed to experimentally isolate blade-level forces, torque, and pitching moment in single-bladed cross-flow turbines.
\item  Experimental results strongly agree with large-eddy simulations, validating the methodology.
\item  The pitching moment, often ignored in blade element momentum models, significantly opposes power generation from the tangential force at high tip-speed ratios.
\item Blade normal forces are up to six times larger than tangential forces, which has implications for structural design.
\end{highlights}

\begin{keyword}
cross-flow turbine \sep vertical-axis turbine \sep current energy \sep pitching moment
\end{keyword}

\end{frontmatter}


\section{Introduction and Background}
Cross-flow turbines, known as vertical-axis wind turbines or ``VAWTs'' in the wind sector, are designed to harness the kinetic energy available in wind, as well as tidal and river currents. These turbines rotate perpendicular to the incoming fluid, and as a result, the blades experience complex, cyclic forcing that varies substantially throughout each rotation. The net mechanical power produced by a cross-flow turbine is the fluid power harnessed by the blades minus the power consumed by support structures (e.g., plates, struts). Individual contributions from these sources can be difficult to isolate through experiments, but doing so can provide fundamental insight into turbine operation. 
Further, numerical models of cross-flow turbines often only include the blades in the simulation domain, either to reduce computational cost for 3D simulation or by necessity for 2D simulation  \cite{Mukul,Mukul2,rezaeiha2018solidity,Coriolis}. Thus, isolating blade-level forces is helpful for validation and comparison between experimental and numerical studies. Additionally, information about the location, magnitude, and direction of the blade resultant force is needed to determine loads on mounting points and internal stresses within the blades. 

Unfortunately, it can be difficult to directly measure blade-level forces and torques in experiments. Le Fouest et al. \cite{sebstalldilema,sebtimescales,mulleners2023pitchcontrol} have measured forces on a laboratory-scale cantilevered blade with a custom array of strain gauges.  McAdam et al. \cite{McAdam} and Bharath et al. \cite{Bharath} have performed similar studies for larger turbines in a laboratory flume and open water site, respectively. Li et al. \cite{LIpressure,LI2016solidity} have measured blade-level forces from pressure tap measurements in wind tunnel tests. However, the majority of cross-flow turbine experiments use simpler load cells that necessitate measuring torque and force at the axis of rotation \cite{hunt2023parametric,RossScaling,Briancontrol,Re,Scherl2020,BachantRef,MillerSolid}. Here, we present a methodology for extracting blade-level forces and torques from experimental measurements at the axis of rotation. Application of this method provides insight into blade-level contributions to turbine performance. 


\subsection{Blade Dynamics}

\begin{figure*}[t!]
    \centering
    \includegraphics[width=1\linewidth]{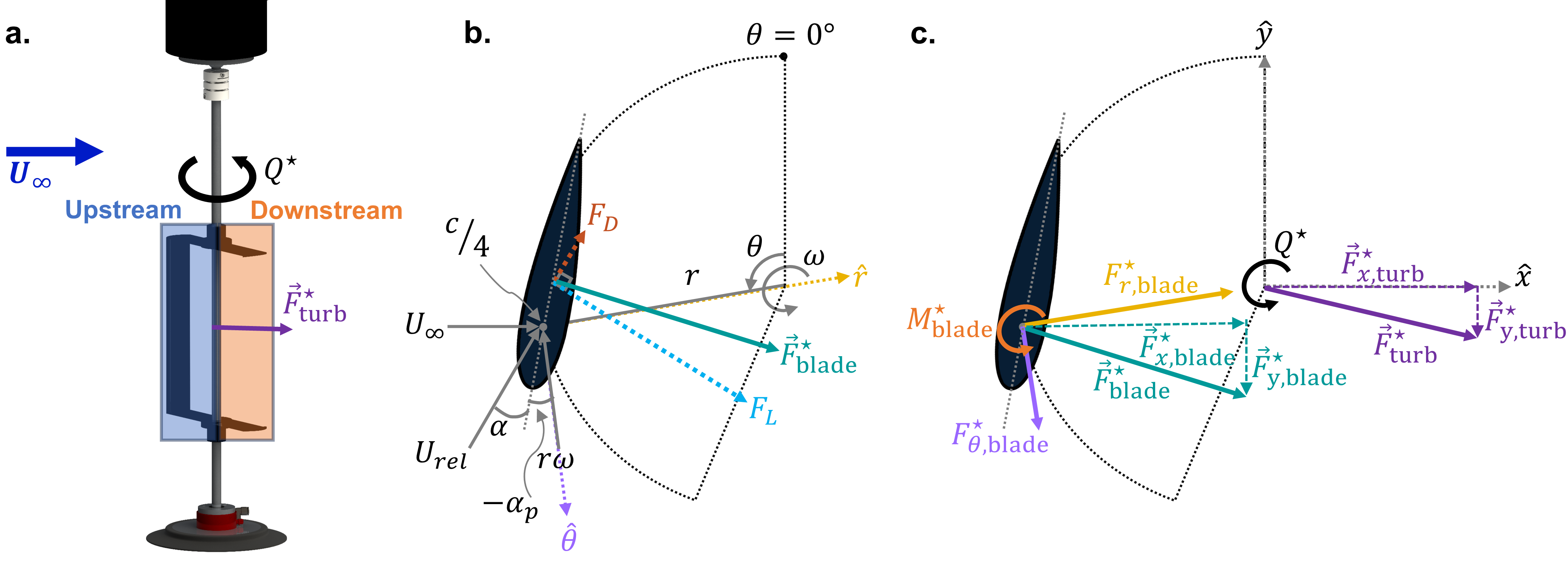}
    \caption{(a) Cross-flow turbine overview with upstream and downstream sweeps indicated. (b) Blade-level kinematic definitions and blade-level coordinate system $(\hat{r},\hat{\theta})$. (c) Corresponding free body diagram and global coordinate system $(\hat{x},\hat{y})$.}
    \label{FBD}
\end{figure*}

At the axis of rotation, cross-flow turbines experience a thrust force in the flow direction, $F_x$, a lateral force perpendicular to the flow direction, $F_y$, and a torque about the axis of rotation, $Q$, that are functions of the turbine's rotation rate. The dimensionless form of the rotation rate is the tip-speed ratio, defined as $\lambda = \frac{r \omega}{U_\infty}$ where $r$ is the turbine radius to the quarter chord, $c/4$, $\omega$ is the rotation rate, and $U_\infty$ is the freestream velocity. These forces and torque vary with the azimuthal position, $\theta$, because, as the turbine rotates, the blades encounter a continually fluctuating relative inflow velocity, $U_{rel}$, and angle of attack, $\alpha$, that affect the direction and magnitude of the blade resultant force (Figure \ref{FBD}a-b). Due to changes in the inflow velocity from momentum extraction, force and torque differ substantially between the first half of the turbine rotation (the ``upstream sweep'') and the second half of the rotation (the ``downstream sweep'') \cite{SnortlandDownstream}. We define $\theta\:=\:0^\circ$ where the blade tangential velocity vector, $r\omega$, is pointed directly upstream. Here, the term ``sweep'' refers to a positional reference in the blade rotation with the upstream sweep corresponding to $0^\circ\:\leq\:\theta\:<\:180^\circ$ and the downstream sweep corresponding to $180^\circ\:\leq\:\theta\:<\:360^\circ$.

At the blade-level, fluid power depends on the tangential projection of the resultant force in the direction of rotation, the pitching moment, $\omega$, and $r$ (Figure \ref{FBD}c). The resultant force, $\vec{F}^\star_{\textrm{blade}}$, is a combination of lift and drag that acts at varying chordwise positions (Figure \ref{FBD}b). Here $^\star$ denotes fluid forces or torques arising from the interaction of the rotating components and inflow (i.e., measured forces less system inertia, Section \ref{fluid force and torques}). Throughout, the subscript ``blade'' refers to the ``blade-level'' contributions which are distinct from ``turbine-level'' measurements (subscript ``turb'') that include contributions from support structures. We assume $\vec{F}^\star_{\textrm{blade}}$ is uniformly distributed along the blade span, and, for convenience, we can prescribe this force to act at $c/4$, by introducing a pitching moment, $M^\star_{\textrm{blade}}$, that accounts for the actual displacement of the center of pressure (Figure \ref{FBD}b,c). As defined in Figure \ref{FBD}, the time-varying normal, $F^\star_{r}$, and tangential, $F^\star_{\theta}$, components of $\vec{F}^\star_{\textrm{blade}}$ are computed as
\begin{equation}
F^\star_{r,\textrm{blade}}(t)\:=\:{F}^\star_{x,\textrm{blade}}(t)\sin\theta(t) - {F}^\star_{y,\textrm{blade}}(t)\cos\theta(t),
\label{normdecomp}
\end{equation}
and
\begin{equation}
F^\star_{\theta,\textrm{blade}}(t)\:=-{F}^\star_{x,\textrm{blade}}(t)\cos\theta(t) - {F}^\star_{y,\textrm{blade}}(t)\sin\theta(t).
\label{tandecomp}
\end{equation}
The time-varying torque produced by a blade is 
\begin{equation}
Q^\star_{\textrm{blade}}(t)\:=\:F^\star_{\theta,\textrm{blade}}(t)r + M^\star_{\textrm{blade}}(t). 
\label{torque}
\end{equation}
Torque is always referenced at the turbine's rotational axis. 

The tangential projection of lift produces torque in the direction of rotation and the tangential projection of drag opposes rotation (Figure \ref{FBD}b). The pitching moment may act with or against rotation depending on its sign, with positive corresponding to a pitch-in moment (towards the center of rotation) acting in the direction of blade rotation (Figure \ref{FBD}c). While the pitching moment about the quarter chord for a symmetric foil in steady flow is zero, for a cross-flow turbine flow curvature \cite{MiglioreWolfe} results in non-zero pitching moments \cite{MukulPOD,sebtimescales}. Flow curvature arises from the variation in the angle of attack and relative velocity along the blade chord such that a physically symmetric foil behaves like a cambered one \cite{MiglioreWolfe}. Non-zero pitching moments are also produced by dynamic stall, an unsteady, non-linear process \cite{Me,Snortland2023,MukulPOD,sebstalldilema,BIANCHINI2016329,Simao,BUCHNER,Dunne} that arises from continuous fluctuations in the angle of attack. Dynamic stall is strongest at lower tip-speed ratios, decreases in severity as the tip-speed ratio increases, and may not occur at high tip-speed ratios if the angle of attack range is sufficiently small \cite{SnortlandDownstream}. 

Several groups have demonstrated the significance of the pitching moment for cross-flow turbines. Using unsteady RANS simulations, Bianchini et al. \cite{BianchiniMpitch} showed that the pitching moment drives asymmetry between the upstream and downstream sweeps during startup. Similarly, in prior work, our group \cite{SnortlandDownstream} hypothesized the pitching moment contributes to performance differences between the upstream and downstream sweep in normal operation. Separately, Mohamed et al. \cite{Mohamed2024} highlighted differences in the pitching moment between a turbine operating in isolation versus as part of an array. Dave et al. \cite{MukulPOD} utilized large eddy simulations to attribute fluctuations in the observed pitching moment to dynamic stall at low tip-speed ratios and to virtual camber at high tip-speed ratios. Finally, Le Fouest et al. \cite{sebtimescales} experimentally identified dynamic stall events for a cross-flow turbine blade using direct measurements of forces and the pitching moment. Despite the acknowledged significance of the pitching moment, this term is often ignored in blade element momentum models, with the exception of 
\cite{Melani2021,Akinori1990}, and descriptions of blade-level forcing. While potentially a consequence of its irrelevance to power production for axial-flow turbines \cite{BianchiniMpitch}, this omission could stem from a historical oversight.  Strickland \cite{StricklandDMST}, who is credited for the first application of a multiple stream tube momentum model for cross-flow turbines, states that, ``the pitching moment on the blade element...is of no consequence for calculation of rotor performance''. A similar assumption carries into Paraschivoiu's development of a double multiple streamtube model \cite{paraschivoiuDMST} and subsequent enhancements (e.g., \cite{BEDON2013,ABDULAKBAR2016,Pucci2022,Beri2011}). In CFD simulations, the pitching moment is implicitly included when torque is computed as the cross-product of the resultant force on the blade and the distance from the center of pressure. If the resultant force is translated to a reference location, such as the quarter chord, a pitching moment arises, but this is not often explicitly reported.

\subsection{Estimating Blade-level Forces and Torques}

\begin{figure*}[t!]
    \centering
    \includegraphics[width=1\linewidth]{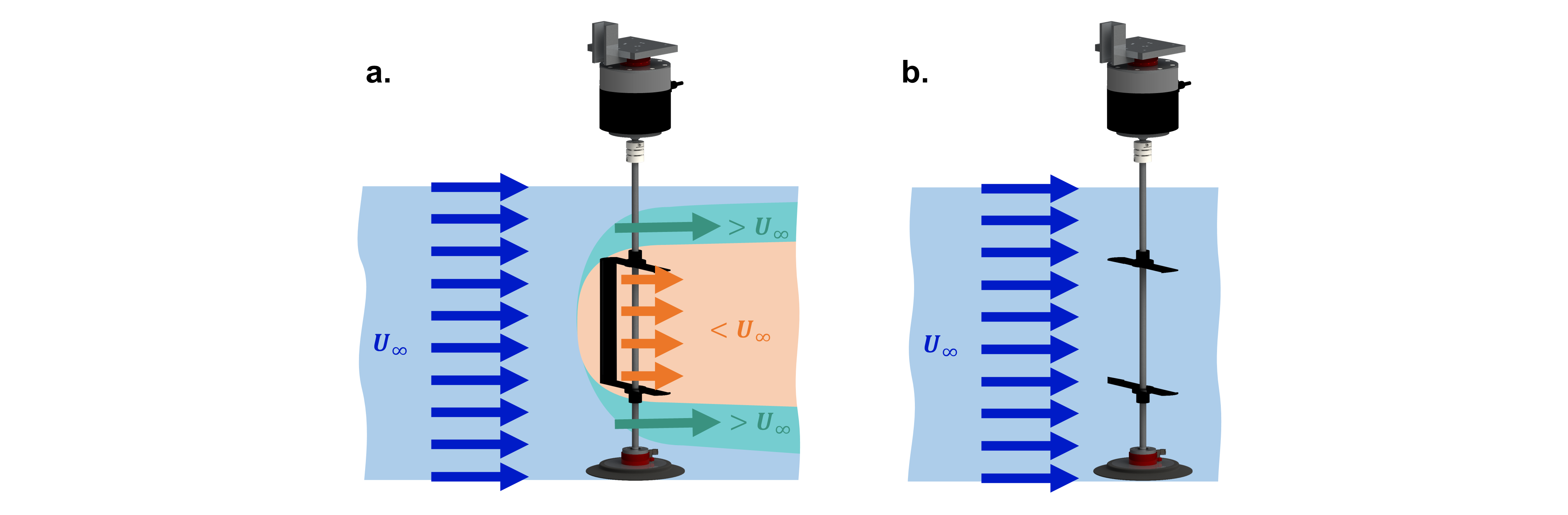}
    \caption{Conceptual representation of how the flow incident on the support structures may differ when (a) blades are present vs. (b) when absent.}
    \label{sups vs turb vel}
\end{figure*}

Superposition is a simple strategy to estimate the blade-level contribution to torque from measurements at the axis of rotation. Inspired by Li and Calisal \cite{LI2010ArmEffects}, Bachant and Wosnik \cite{BachantRef}, and Bianchini et al. \cite{bianchini2012improved}, Strom et. al. \cite{stromsupports} demonstrated that blade-level torque can be approximated by subtracting the parasitic torque measured in experiments with only the support structure present from torque measurements of the full turbine at the same operating condition. 
This superposition principle relies on the assumptions that (1) the torque measured with the blades removed is representative of the oppositional torque from the support structures in a full turbine and (2) secondary interactions between the blades and support structures are minimal. Neither of these assumptions are fully realized in practice. 
The presence of the turbine substantially modifies the incoming flow such that flow is decelerated through the turbine and accelerated around it \cite{SnortlandDownstream}. While unknown \emph{a priori}, this induction may cause torque consumed by support structures to vary substantially inside versus outside the turbine span, as well as differences when blades are present versus removed (Figure \ref{sups vs turb vel}b). 
As a result, assumption (1) is clearly violated for the central shaft, but shaft forces negligibly contribute to the torque balance due to the small moment arm between the axis of rotation and shaft surface.
For similar reasons assumptions (1) and (2) are also violated for blade-shaft connections (such as struts or disks), which consume substantial torque.  Despite this, Strom et. al. \cite{stromsupports} demonstrated that the torque superposition strategy reasonably collapsed blade-level power coefficients for a range of blade-shaft connection types. 
The residual deviations between connection types are likely due to induction and/or secondary interactions. If an analogous superposition approach was effective for forces measured at the axis of rotation, it would be straightforward to estimate blade-level normal and tangential forces, as well as the pitching moment. However, 
forces on the central shaft are appreciable, are anticipated to be the largest contribution to support structure force (verified in Section \ref{results}), and are expected to differ substantially with changes to induction when blades are present. 

This work explores the efficacy of blade-level force, torque, and pitching moment identification from turbine-level measurements at the axis of rotation for a single-bladed turbine. To do so, we conduct experiments with four shaft configurations to verify a physics-based superposition strategy. 
The accuracy of the derived blade-level values are then assessed through comparison to equivalent blade-only, large-eddy simulations. This paper is structured as follows. Section \ref{methods} lays out the experimental setup, the procedure for identifying the blade-level values, and the simulation methods. In Section \ref{results}, we compare experimental forces, torque, and the pitching moment with simulation. In Section \ref{discussion}, we discuss the efficacy of the method and demonstrate the importance of the pitching moment on turbine performance. 

\section{Methods}
\label{methods}

\subsection{Determination of Fluid Forces and Torques}
\label{fluid force and torques}
The first step to estimating blade-level quantities is to isolate the forces and torques arising solely from fluid-structure interactions in the experimental measurements. Newton's second law describes the sum of the forces and torques acting on the turbine at the center of rotation as 
\begin{equation}
\sum \vec{F}\:=\:m\vec{a_g}
\end{equation}
and
\begin{equation}
\sum Q\:=\:J\dot{\omega}
\end{equation}
where $m$ is the mass of the rotating system, $\vec{a_g}$ is the acceleration of the center of mass, and $J$ is the rotational moment of inertia. From this, the fluid torques, $Q^\star$, and forces, $\vec{F}^\star$, acting on the turbine elements (i.e., full turbine, support structures, shaft) are related to measured values as
\begin{equation}
\label{F hydro}
\vec{F}^\star(t)\:=\:m\vec{a_{g}}(t) + \vec{F}_\textrm{m}(t)
\end{equation}
and
\begin{equation}
\label{Q hydro}
Q^\star(t)\:=\:J\dot{\omega}(t) + Q_\textrm{m}(t)
\end{equation}
where the subscript ``m'' denotes the sum of the forces and torque measured by the pair of load cells in the global coordinate frame (Figure~\ref{FBD}).
For solid body rotation, the components of $\vec{a_g}$ are $r_g\omega^2$ in the radial direction and $r_g\dot{\omega}$ in the tangential direction where $r_g$ is the radius to the center of mass \cite{MachinesMechText}. These components are expressed in the Cartesian coordinates as 
\begin{equation}
a_{g,x}(t)\:=\:r_g[\omega(t)^2\sin(\theta(t)-\delta\theta)-\dot{\omega}(t)\cos(\theta(t)-\delta\theta)],
\end{equation}
and
\begin{equation}
a_{g,y}(t)\:=\:r_g[-\omega(t)^2\cos(\theta(t)-\delta\theta)-\dot{\omega}(t)\sin(\theta(t)-\delta\theta)]
\end{equation}
where $\delta\theta$ is the phase offset of the center of mass from the quarter chord. 
Under constant speed control, $\dot{\omega}$ is negligible, so after substituting and simplifying, fluid forces and torque are expressed as
\begin{equation}
F^\star_{x}(t)\:=\:mr_g\omega^2\sin(\theta(t)-\delta\theta) + F_{x,\textrm{m}}(t),
\label{hydroX}
\end{equation}
\begin{equation}
F^\star_{y}(t)\:=\:-mr_g\omega^2\cos(\theta(t)-\delta\theta) + F_{y,\textrm{m}}(t),
\label{hydroY}
\end{equation}
and
\begin{equation}
Q^\star(t)\:=\:Q_\textrm{m}(t).
\label{hydroQ}
\end{equation}
The product of the rotating mass and radius to the center of mass, $mr_g$, and $\delta\theta$ are determined by conducting experiments in quiescent air (i.e., with negligible fluid forcing) as described in Appendix \ref{inertia}. Both terms are negligible for the axisymmetric experiments (i.e., support structure and shaft-only). Since the time integrals of $\cos\theta(t)$ and $\sin\theta(t)$ are zero, the time-averaged fluid forces are equivalent to the time-averaged measured quantities at the axis of rotation, regardless of rotational symmetry. However, the distinction between fluid and measured quantities is important for phase-averaged representations.

\subsection{Non-dimensional Coefficients}

Because turbine forcing is a function of the tip-speed ratio and azimuthal blade position, 
it is convenient to define non-dimensional force, torque, and performance coefficients on a phase-average basis, denoted by $\langle X \rangle$. For turbines rotating at a constant speed, time averages, denoted by $\overline X$, are equal to the azimuthal average of $\langle X \rangle$. 
Any force component at the turbine- or blade-level, such as $F^\star_x$ $F^\star_y$, $F^\star_r$, or $F^\star_\theta$, may be non-dimensionalized as
\begin{equation}
C_{F,i}(\lambda,\theta)\:=\:\frac{ \langle F_i^\star(\lambda,\theta) \rangle}{\frac{1}{2}\rho {U^2_\infty} DS},
\label{forcenorm}
\end{equation} 
where $D$ is the turbine diameter based on the outer blade surface and $S$ is the blade-span.
We choose to non-dimensionalize the pitching moment, in a manner akin to aerodynamic coefficients, as
\begin{equation}
C_{M}(\lambda,\theta)\:=\:\frac{ \langle M^\star(\lambda,\theta) \rangle}{\frac{1}{2}\rho (r\omega)^2 Sc^2}.
\label{pitchnorm}
\end{equation} 
The tangential velocity is used instead of the relative velocity in Equation \ref{pitchnorm}, as the latter changes with blade position and is unknown in these experiments. Torque is non-dimensionalized as
\begin{equation}
C_{Q}(\lambda,\theta)\:=\:\frac{ \langle Q^\star(\lambda,\theta) \rangle}{\frac{1}{2}\rho {U^2_\infty} DSr},
\label{CQ}
\end{equation} 
and the performance coefficient is defined as
\begin{equation}
C_{P}(\lambda,\theta)\:=\:\frac{\langle Q^\star(\lambda,\theta) \omega \rangle}{\frac{1}{2}\rho \overline{U^3_\infty} DS} \:=\:C_{Q}(\lambda,\theta)\lambda.
\label{CP}
\end{equation}
Throughout, we refer to the performance coefficient using the shorthand of ``performance''. 

From Equations \ref{normdecomp}, \ref{tandecomp}, and \ref{forcenorm}, the normal and tangential force coefficients may be expressed in terms of streamwise and cross-stream force coefficients as 
\begin{equation}
C_{F,r}(\lambda,\theta)\:=\:C_{F,x}\sin\theta - C_{F,y}\cos\theta
\label{norm from coeffs}
\end{equation}
and
\begin{equation}
C_{F,\theta}(\lambda,\theta)\:=\:-C_{F,x}\cos\theta - C_{F,y}\sin\theta.
\label{tan from coeffs}
\end{equation}
The torque coefficient at the center of rotation is a function of the tangential force and pitching moment coefficients as
\begin{equation}
C_{Q}(\lambda,\theta)\:=\:C_{F,\theta} + C_{M}\lambda^2\bigg(\frac{c^2}{Dr}\bigg).
\label{torque from coeffs}
\end{equation}
In Equation \ref{torque from coeffs}, the $\lambda^2\frac{c^2}{Dr}$ term arises due to the different velocity and length scales in non-dimensionalization of each coefficient (Equations \ref{forcenorm} and \ref{pitchnorm}).
In summary, $C_Q$ is computed from the fluid torque (Equation \ref{CQ}) and  $C_{F,r}$ and $C_{F,\theta}$ are computed from the fluid force coefficients (Equations \ref{norm from coeffs} and \ref{tan from coeffs}). From these, the pitching moment coefficient can be found by re-arrangement of Equation \ref{torque from coeffs}. 

\subsection{Determination of Blade-level Forces and Torque}
\label{to blade}

\begin{figure*}[t!]
    \centering
    \includegraphics[width=1\linewidth]{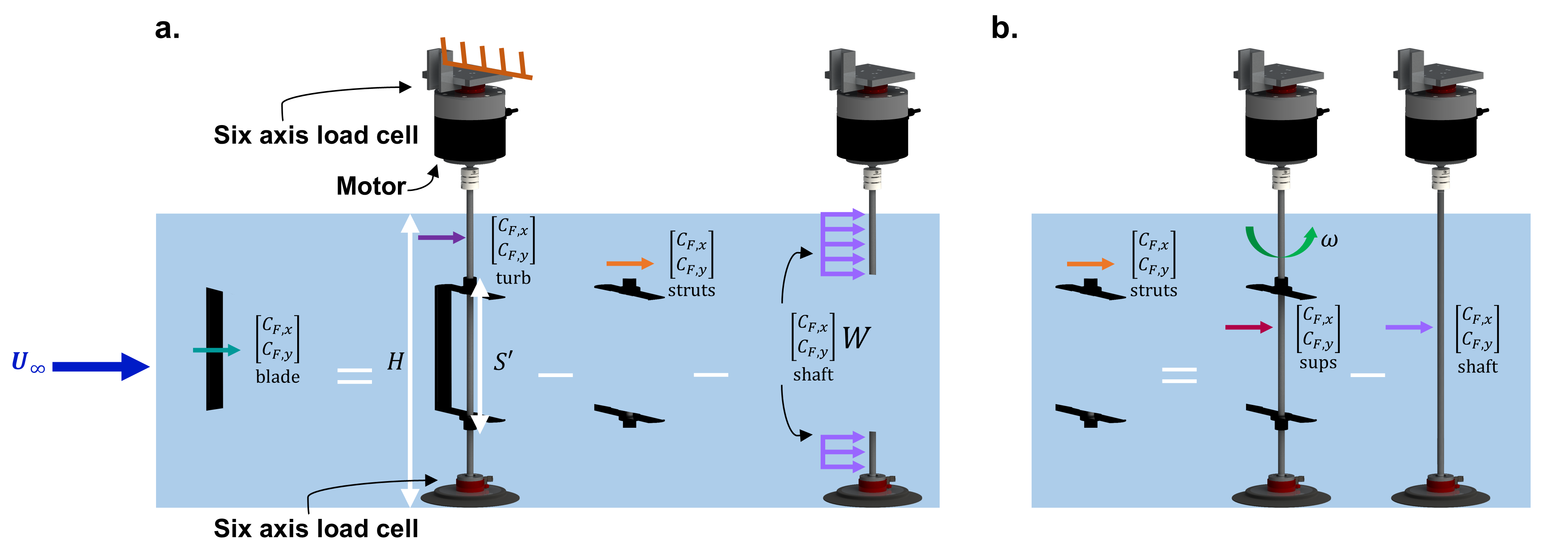}
    \caption{(a) Schematic of isolation of the blade-level forcing using the physics-based strategy.  (b) Schematic of isolation of forcing on the struts.}
    \label{substratpipeline}
\end{figure*}

Figure \ref{substratpipeline} provides an overview of the proposed physics-based strategy for isolating blade-level forces and torque from turbine-level measurements. 
Torque and force contributions from each blade are ambiguous in a multi-bladed turbine, so this method is only applicable to single-bladed turbines. The support structure influence is modeled as a superposition of the forces and torque acting on different turbine components, inspired by Strom et al. \cite{stromsupports}, with adjustments that account for induction and other flow physics. This approach is performed on a non-dimensional basis using phase-average force and torque coefficients for the turbine elements (i.e., full turbine, struts and shaft, shaft only) measured in separate experiments at equivalent tip-speed ratios. Additional subtleties of the implementation are elucidated in Appendix \ref{superpos deets}.

The shaft is a rotating cylinder that experiences a lift force perpendicular to the inflow (Magnus effect) and a drag force parallel to the inflow. These forces depend on the Reynolds number and the shaft's non-dimensional rotation rate, $k$ (ratio of the shaft surface speed to the inflow). Snortland et al. \cite{SnortlandDownstream} showed that induction in the upstream sweep results in low incident velocities on the shaft inside the turbine span (i.e., reduced shaft Reynolds numbers, increased $k$). The drag coefficient on a rotating cylinder decreases as the Reynolds number decreases and has a moderate dependency on $k$. On the other hand, the lift coefficient strongly increases with $k$ and is less sensitive to the Reynolds number \cite{Ma2022cylinder}. Guided by these physics, we model the shaft contribution to the turbine-level thrust force (produced by drag) differently than the contribution to the lateral force (produced by lift). Specifically, within the turbine span, the thrust contribution from the shaft is assumed to be negligible, while the lateral force contribution is preserved. 

As depicted in Figure \ref{substratpipeline}a, the blade-level thrust and lateral force coefficients are computed as
\begin{equation}
\label{blade level coeffs}
[C_{F,x}, C_{F,y}]_{\textrm{blade}} \:=\:[C_{F,x}, C_{F,y}]_{\textrm{turb}}  -  [C_{F,x}, C_{F,y}]_{\textrm{struts}}  -  [C_{F,x}, C_{F,y}]_{\textrm{shaft}} \cdot \: \boldsymbol{W},
\end{equation}
where the subscript ``struts'' denotes contributions from the struts only. Blade-level torque is computed in the same manner as Strom et al. \citep{stromsupports} as
\begin{equation}
\label{blade level coeffs torque}
 C_{Q,\textrm{blade}} \:=\: C_{Q,\textrm{turb}}  -   C_{Q,\textrm{sups}}. 
\end{equation}
where the subscript ``sups'' denotes contributions from the combination of struts and shaft.
From these values, the blade-level normal, tangential, and pitching moment coefficients are computed via Equations \ref{norm from coeffs} - \ref{torque from coeffs}.

The contribution from the struts is isolated by subtracting shaft-only coefficients, $[C_{F,x}, C_{F,y}]_{\textrm{shaft}}$, from coefficients with the entire support structure present, $[C_{F,x}, C_{F,y}]_{\textrm{sups}}$ (i.e., $[C_{F,x}, C_{F,y}]_{\textrm{struts}}\:=\:[C_{F,x}, C_{F,y}]_{\textrm{sups}}-[C_{F,x}, C_{F,y}]_{\textrm{shaft}}$, Figure \ref{substratpipeline}b). Then, the contribution from the shaft to turbine-level forces is weighted by a factor $\boldsymbol{W}$ that reflects the limited thrust force on the shaft inside the blade span. For the thrust component, $\boldsymbol{W_x}\:=\:\frac{H_\Psi-S'}{H_\Upsilon}$ represents the proportion of the thrust on the shaft outside of the turbine span. This formulation accounts for small differences in $H$, mean dyanmic water depth, for when the blades are present versus when they are absent (Appendix \ref{superpos deets}). $S'$ represents the total turbine span (blade + struts + couplings, 25.2 cm) and the subscripts $\Psi$ and $\Upsilon$ denote the full-turbine and shaft-only conditions, respectively. For the lateral force, shaft loading is scaled only by the small differences in $H$ ($\boldsymbol{W_y}\:=\:\frac{H_\Psi}{H_\Upsilon} \sim 1$).  

\subsection{Experimental Setup and Data Processing}
\label{flumedescript}
\begin{figure*}[t!]
    \centering
    \includegraphics[width=1\linewidth]{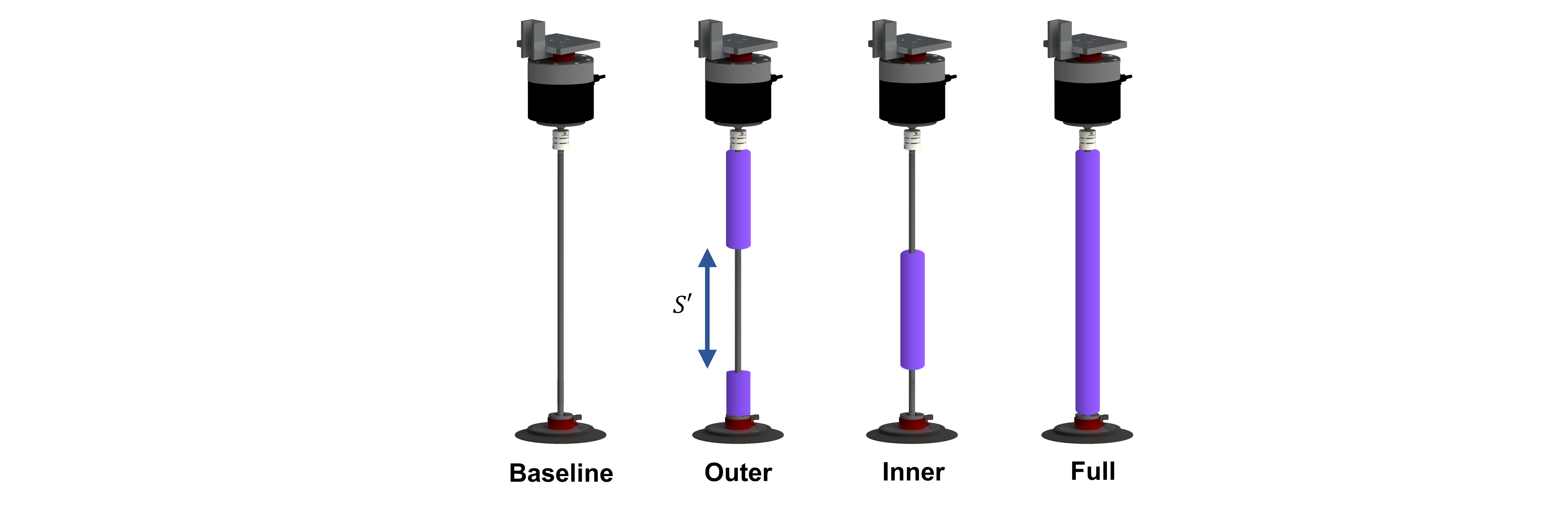}
    \caption{Schematics of the four different turbine shaft configurations tested: (a) ``Baseline'', (b) ``Outer'', (c) ``Inner'', and (d) ``Full''.}
    \label{shafts}
\end{figure*}

Experiments to test this approach were performed in the Alice C. Tyler flume at the University of Washington. MATLAB Simulink Desktop Real-Time was used for data collection and turbine control. Turbine torque and forces for a one-bladed turbine were measured at the center shaft. The test setup utilized a Yaskawa SGMCS-05B servomotor rigidly coupled to the flume cross beam and controlled with a Yaskawa SGDV-2R1F01A002 drive. Forces and torque were measured by a pair of six-axis reaction load cells (ATI Mini45) at the mount points for the servomotor and bottom bearing. The servomotor regulated the rotation rate to achieve a desired tip-speed ratio for a given test. The servomotor encoder measured the blade position with a resolution of $2^{18}$ counts/rotation and $\omega$ was computed by differentiation. For each control set point (i.e., a single $\lambda$), data were acquired for $60$ seconds at 1 kHz, and the standard deviation in rotation rate was 0.1\% of the mean rotation rate on average. 

The single-bladed turbine consisted of a NACA 0018 foil blade with NACA 0008 foil struts supporting each end of the blade span. The struts are designed for two-bladed tests and, as such, are axisymmetric. Appendix \ref{superpos deets} includes comparable results for a turbine with disk end-plates. The turbine had a radius of 8.2 cm to $c/4$, a diameter based on the outer blade surface of 17.2 cm, a blade-span of 23.4 cm, a chord length of 4.06 cm, and a $-6^\circ$ (toe-out) preset pitch angle. To produce different shaft forces, four shaft configurations with differing diameters were tested (Figure \ref{shafts}). The ``Baseline'' configuration consisted of a 1.27 cm diameter shaft and the other three configurations utilized 3.81 cm diameter collars fit over the Baseline shaft. The collars were placed outside of the turbine span for the ``Outer'' configuration, inside the turbine span for the ``Inner'' configuration, and along the entire shaft for the ``Full'' configuration. In Equation \ref{blade level coeffs}, $[C_{F,x}]_{\textrm{ shaft}}$ corresponds to the shaft type outside the turbine swept area. For example, the ``Baseline'' shaft-only data is used to obtain blade-level thrust for both the ``Baseline'' and ``Inner'' configurations. However, for blade-level lateral force and torque no such adjustment is made (i.e. ``Inner'' shaft-only data is used for the ``Inner'' configuration).

The flume is $0.75$ m wide and experimental conditions (Table \ref{exp params}) were adjusted to keep the depth-based Froude number, $Fr\:=\:U_\infty/\sqrt{gH}\:=\:0.37$, blockage ratio, $\beta\:=\:\frac{A_F}{A_C}\:=\:12.7\%$, and Reynolds number, $Re_c\:=\: \frac{U_{\infty}c}{\nu}\:=\:4.5\times10^4$, constant across the four shaft configurations. Here $g$ is the gravitational constant and $A_F$ is the frontal area of the turbine and all submerged components, $A_C$ is the channel cross-sectional area, and $\nu$ is the kinematic viscosity (function of temperature). $U_{\infty}$ was measured by an acoustic Doppler velocimeter (Nortek Vectrino) sampling at 16 Hz and positioned $5D$ upstream of the turbine. The turbulence intensity was 1-2\% for all tests. The $(U_\infty)^2$ and $(U_\infty)^3$ terms in Equations \ref{forcenorm}, \ref{CQ}, and \ref{CP} are computed as a time-average of all of the squared or cubed freestream velocity measurements acquired at a tip-speed ratio set point, respectively. This approach assumes azimuthal variations in force and torque are significantly larger than inflow fluctuations, and measurement noise remains minimal.  While appropriate for these experimental conditions, this would likely be inaccurate in situations with higher turbulence intensity (e.g., field sites \citep{thomson2012}). A free surface transducer measured the dynamic water depth $6.5D$ upstream of the turbine. The water temperature was maintained within $\pm 0.2 ^\circ C$ with a pool heater. Experimental conditions were held constant between tests with the full turbine and corresponding support structure and shaft-only tests. 

\begin{table}
\centering
\caption{Experimental parameters}
\label{exp params}
\begin{tabular}{ l|c|c|c|c }
\toprule
Parameter & ``Baseline'' & ``Outer'' & ``Inner'' & ``Full''\\
\midrule
$A_F$ (m\textsuperscript{2}) & 0.047 & 0.054 & 0.047 & 0.054 \\ 
$H$ (m) & 0.49 & 0.57 & 0.49 & 0.57 \\
$U_\infty$ (m/s) & 0.8 & 0.86 & 0.8 & 0.86 \\
Temperature ($^\circ$C) & 35 & 31.2 & 35 & 31.2 \\
$\nu$ (m\textsuperscript{2}/s) & $7.81\times10^{-7}$ & $7.24\times10^{-7}$ & $7.81\times10^{-7}$ & $7.24\times10^{-7}$ \\
\bottomrule \end{tabular}
\end{table} 

\begin{figure*}[t!]
    \centering
    \includegraphics[width=1\linewidth]{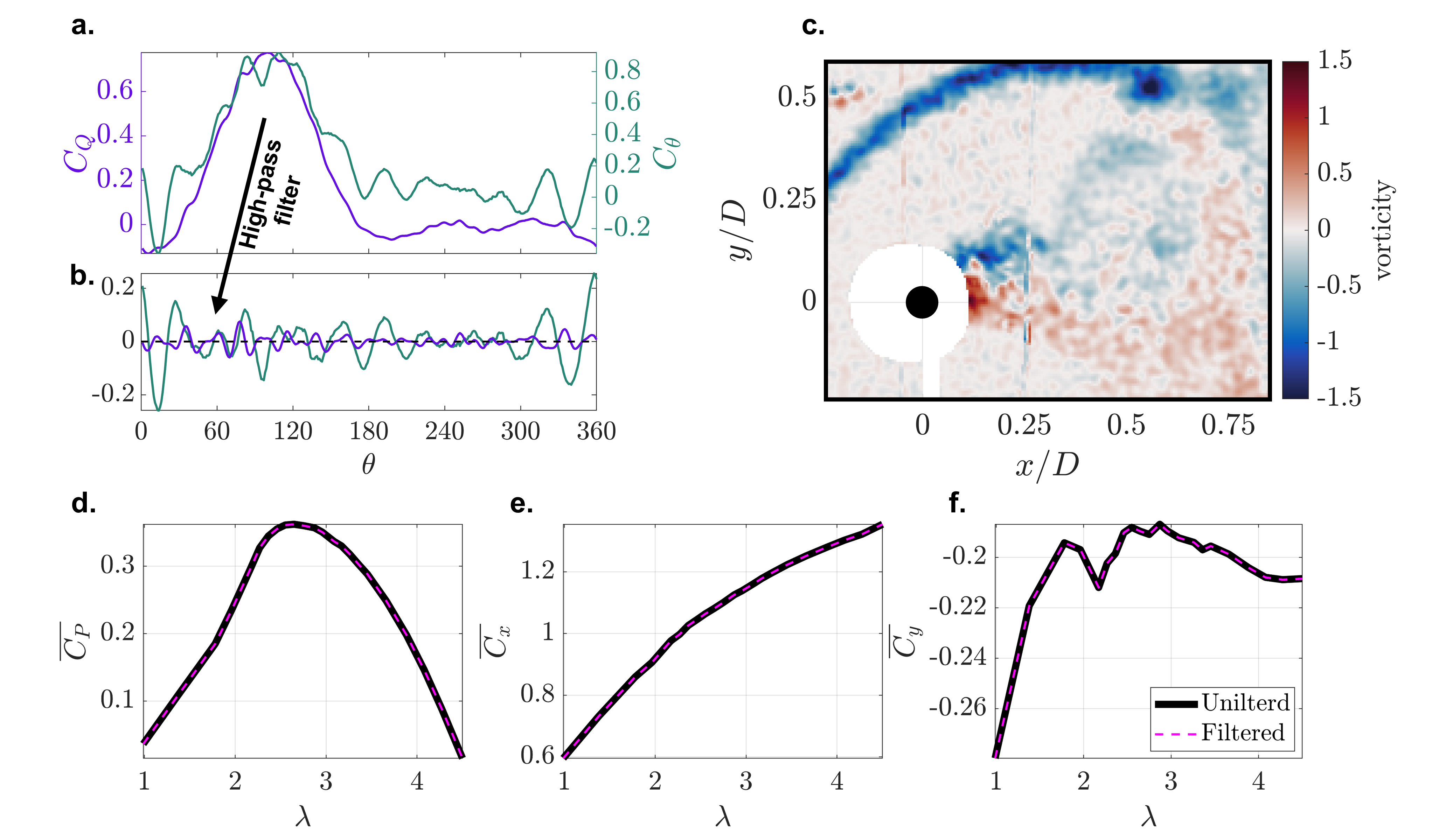}
    \caption{(a) Phase-averaged, turbine-level tangential force and torque coefficients for $\lambda\:=\:2.4$ (b) Phase-averaged secondary tangential force oscillations after high-pass filtering. (c) Phase-averaged vortex shedding off of the central turbine shaft at $\lambda\:=\:2.4, \theta\:=\:201^\circ$. Vorticity field is computed from PIV data from \cite{SnortlandDownstream}. Comparison between filtered and unfiltered, time-averaged (d) performance (e) thrust force, and (f) lateral force coefficients }
    \label{wiggle}
\end{figure*}

All experimental data were filtered with a low-pass, zero-phase, Butterworth filter with a 75 Hz cutoff to remove higher-frequency electromagnetic interference. However, oscillations in the turbine-level tangential force coefficient at frequencies higher than the blade passage remained (Figure \ref{wiggle}a). These oscillations were not present in the filtered torque coefficient, so they are unlikely related to fluid forcing on the blade. However, these could arise from vortex shedding off the turbine shaft (Figure \ref{wiggle}c) and/or a fluid-structure coupling induced by the turbine forcing. Application of an additional high-pass filter with an 10 Hz cutoff frequency produced a time series dominated by the oscillations (Figure \ref{wiggle}b). Consequently, an equivalent low-pass filter was used to remove these oscillations from all force and torque time series. Because the oscillations had near-zero time-averages, this filtering had a negligible effect on time-averaged quantities (Figure \ref{wiggle}d-f). 

\subsection{Simulation Methods}

\begin{figure}[t!]
    \centering
    \includegraphics[width=1\linewidth]{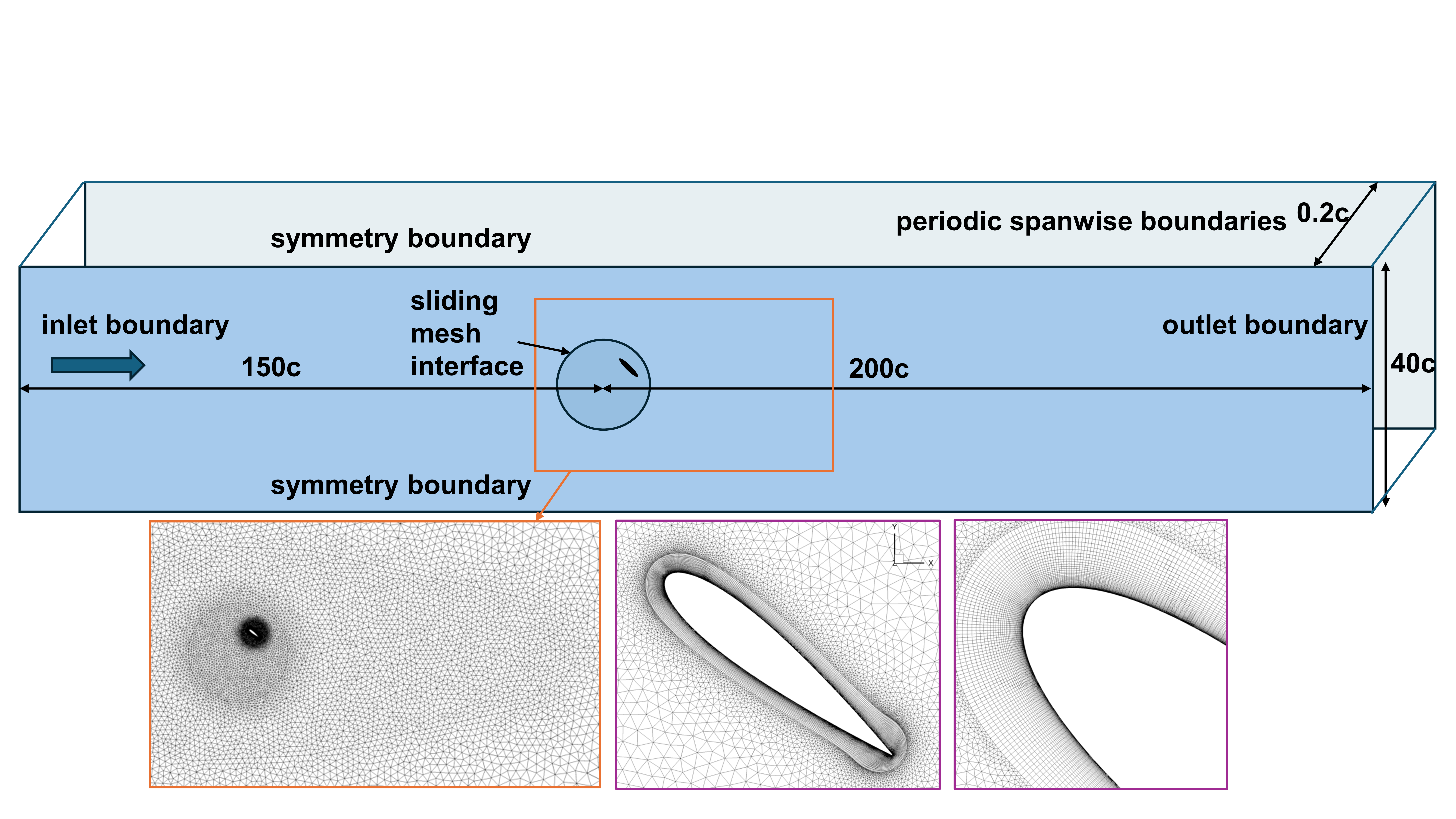}
    \caption{Computational domain, boundary conditions, and images of the mesh resolution close to the blade. }
    \label{fig:cfdmethods}
\end{figure}

To evaluate the accuracy of the physics-based strategy, a large-eddy simulation (LES) of a single blade (no struts or shaft) was performed at conditions matching the experiments. Specifically, the simulations consisted of a single NACA 0018 blade mounted at $c/4$ with $-6^\circ$ preset pitch and a $c/r$ of 0.5.

The simulations solved the low-pass, spatially filtered Navier-Stokes Equations, with a local dynamic $k$-Equation sub-grid scale (SGS) closure model. The computational methodology was a second-order finite volume scheme implemented within the OpenFOAM libraries. Time-stepping was performed with a second-order backwards scheme in increments limited by the Courant-Friedrichs–Lewy (CFL) condition of $\textrm{CFL}<1$, producing high-fidelity time-accurate flows.

The blade was rotated by means of an arbitrary mesh interface (AMI), or a slip boundary, between the rotating mesh of the blade and the outer mesh (Figure ~\ref{fig:cfdmethods}). The inflow had a uniform velocity profile at the inlet and a constant pressure at the outlet. The three-dimensional domain was periodic in the spanwise direction with a computational depth of $0.2c$. Slip walls were implemented on edges of the domain (parallel to the freestream flow direction) that impose an equivalent 2D blockage with respect to the width of the computational domain, $W$, such that $D/W=0.11$. 

The mesh was body-fitted and structured close to the blade with $y^+<1$ on the suction surface of a non-rotating mesh. Outside the boundary layer region, the mesh transitioned to an unstructured format to reduce computational costs. The computational methodology closely follows that previously reported \cite{Mukul2, MukulPOD}. Mesh independence, SGS model sensitivity, and validation against turbine-level experimental performance and flow-fields (PIV) are described in \cite{Mukul2}.

The simulations had a Reynolds number of $Re_c=45,000$ and were performed at $\lambda\:=\:1.4 - 4.3$ by increasing the rotational speed of the blade region and holding all other parameters constant. The blade forces were computed in the frame of reference of the computational domain ($x,y$) by integrating the normal and viscous stresses over the blade profile at every time-step. These were converted to normal and tangential components through a coordinate transformation based on blade orientation. The torque coefficient was computed by the cross-product of the resultant force on the blade and the distance from the center of pressure to the rotational axis. As for experiments, the tangential and normal force coefficients were computed via Equations \ref{norm from coeffs} and \ref{tan from coeffs} and the pitching moment coefficient was obtained from Equation \ref{torque from coeffs}. 

\begin{figure}[t!]
    \centering
    \includegraphics[width=1\linewidth]{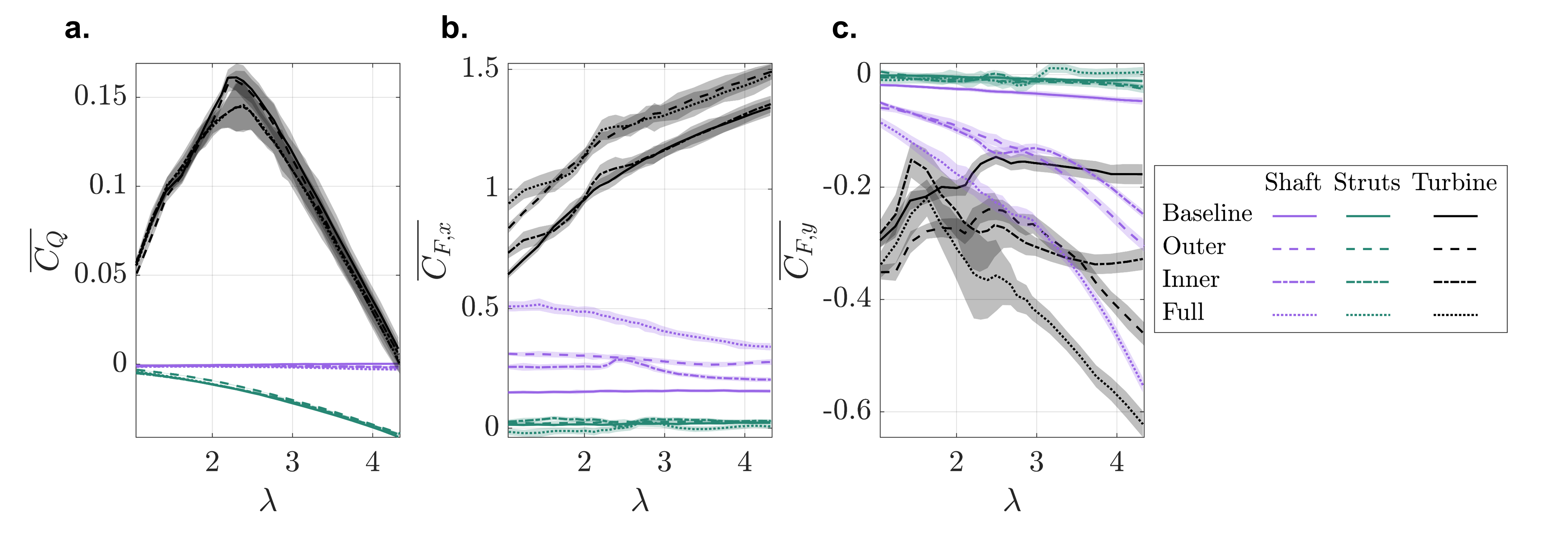}
    \caption{Comparison between experimental time-averaged (a) performance (b) thrust force, and (c) lateral force coefficients for the shaft, struts, and full turbine. The shaded areas denote the standard deviation of time-average quantities for all cycles at each $\lambda$.}
    \label{RAW time avg}
\end{figure}

\section{Results}
\label{results}

Experimental time-averaged torque and force for the turbine, struts, and shaft for the four shaft configurations are presented in Figure \ref{RAW time avg}. As expected, the shaft minimally influences $\overline{C_Q}$, while the struts have a significant effect (Figure \ref{RAW time avg}a). 
There are no appreciable interactions between the shaft and struts (i.e., changing the shaft size does not appreciably affect strut torque or force). The thrust and lateral forces on the shaft are the largest contributors to force on the support structure and depend on the shaft configuration (Figure \ref{RAW time avg}b,c). 
In agreement with canonical results \cite{Ma2022cylinder}, as the tip-speed ratio increases, the thrust force coefficient for the shaft decreases and the magnitude of the lateral force coefficient increases (Section~\ref{to blade}). Shaft configurations with the same diameter outside of the blade span (e.g., ``Baseline'' and ``Inner''), have comparable turbine-level thrust despite a 200\% increase in the shaft diameter inside the blade span for the ``Inner'' configuration. This supports the assumption that the thrust force on the shaft within the blade span is relatively inconsequential. 

\subsection{Comparison of the Torque, Thrust, and Lateral Force Coefficients}
\begin{figure}[t!]
    \centering
    \includegraphics[width=1\linewidth]{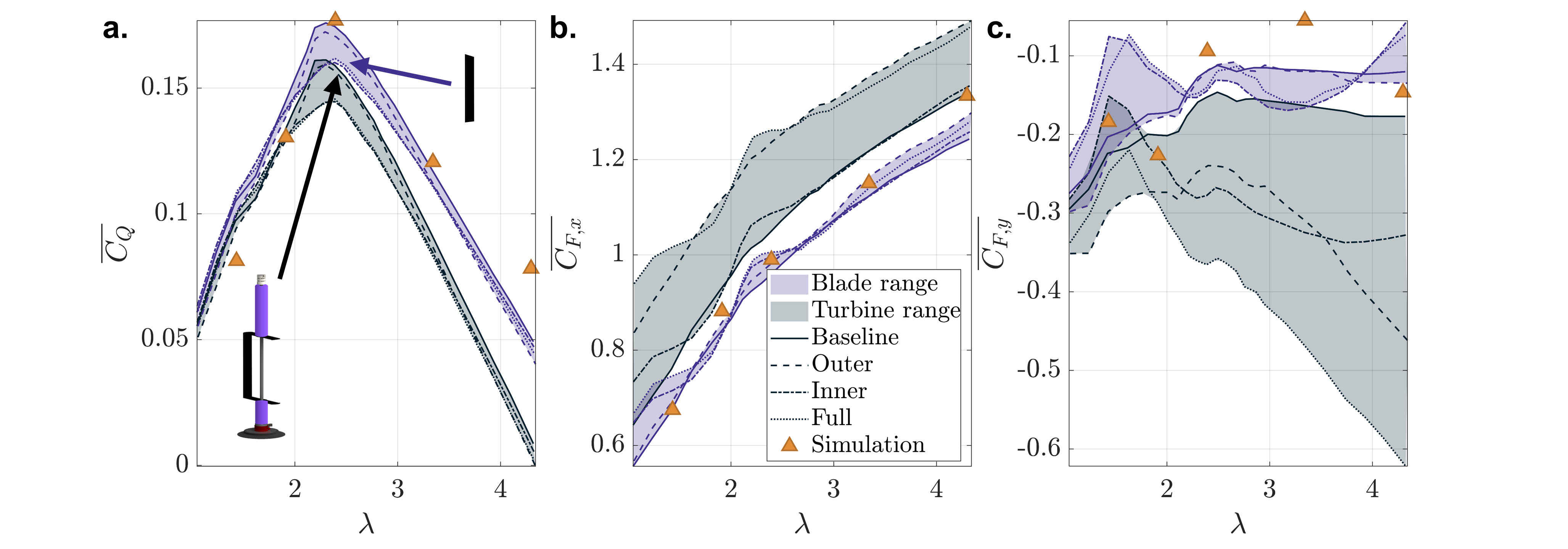}
    \caption{Comparison of full turbine and blade-only contributions to torque and forces. Time-averaged (a) torque (b) thrust force, and (c) lateral force coefficients for the full turbine, the blades, and the simulations. The spread of the colored area corresponds to the range of time-averaged values for the different shaft configurations (Figure \ref{shafts}).}
    \label{Substrat time avg}
\end{figure}

Because the blade-level forces should be approximately invariant with shaft configuration, the differences in turbine-level forces provide a test case for the proposed physics-based strategy. Experimental time-averaged torque and forces at the turbine and blade level are compared to the simulation results in Figure \ref{Substrat time avg}. 
Blade-level torques are elevated relative to turbine-level torques once the parasitic support structure losses are accounted for, while thrusts decrease and lateral forces are less negative when the shaft contributions are accounted for.  
Additionally, consistent with superposition principles, the spread (shaded areas) between the different shaft configurations is substantially reduced for the blade-level thrust and lateral force coefficients relative to the turbine-level results. Despite induction differences and potential secondary interactions between the turbine components neglected in the superposition, we observe nearly complete convergence for blade-level thrust across all shaft configurations. The remaining spread in blade-level thrust is comparable to the standard deviation of time-averaged performance between cycles (Figure \ref{RAW time avg}a,b). Time-averaged thrust is always positive, increases with tip-speed ratio, and is significantly larger than the lateral force. The time-averaged lateral force is always negative and, at the blade level, is less sensitive to the tip-speed ratio than thrust. Across the board, the blade-only simulations have substantially better agreement with experimental blade-level coefficients than with turbine-level coefficients. This agreement is excellent for the thrust coefficient where the simulation values are within the spread of the different shaft configurations except for $\lambda\:=\:4.3$. Comparatively, the torque and lateral force coefficients differ more, but the magnitudes and trends agree reasonably well. 

Phase-averaged coefficients from experiments and simulations are compared in Figure \ref{phase averages} for five tip-speed ratios. Differences between the experimental turbine- and blade-level coefficients appear more subtle in the phase averages. This is partially a consequence of axis scale, as the range of values throughout the cycle masks differences in the phase averages that are on the order of those in the time averages. Thrust is positive throughout the rotation (except for in the downstream sweep at $\lambda\geq 3.3$) and the lateral force oscillates about zero. While the lateral force on the shaft is a large contribution to the turbine-level time-average lateral force (Figures \ref{RAW time avg} and \ref{Substrat time avg}), the phase-average lateral force is dominated by the blades. In general, the blade-level and the simulation values are well matched for all coefficients, but differences are more pronounced in the phase averages than the time averages (e.g., $C_Q$ and $C_{F,y}$ at $\lambda = 1.9$). For these cases, the time-averaged agreement between simulations and the experiments is a consequence of offsetting differences throughout the rotation. 

\subsection{Comparison of the Tangential Force and Pitching Moment Coefficients}
\begin{figure*}[t!]
    \centering
    \includegraphics[width=1\linewidth]{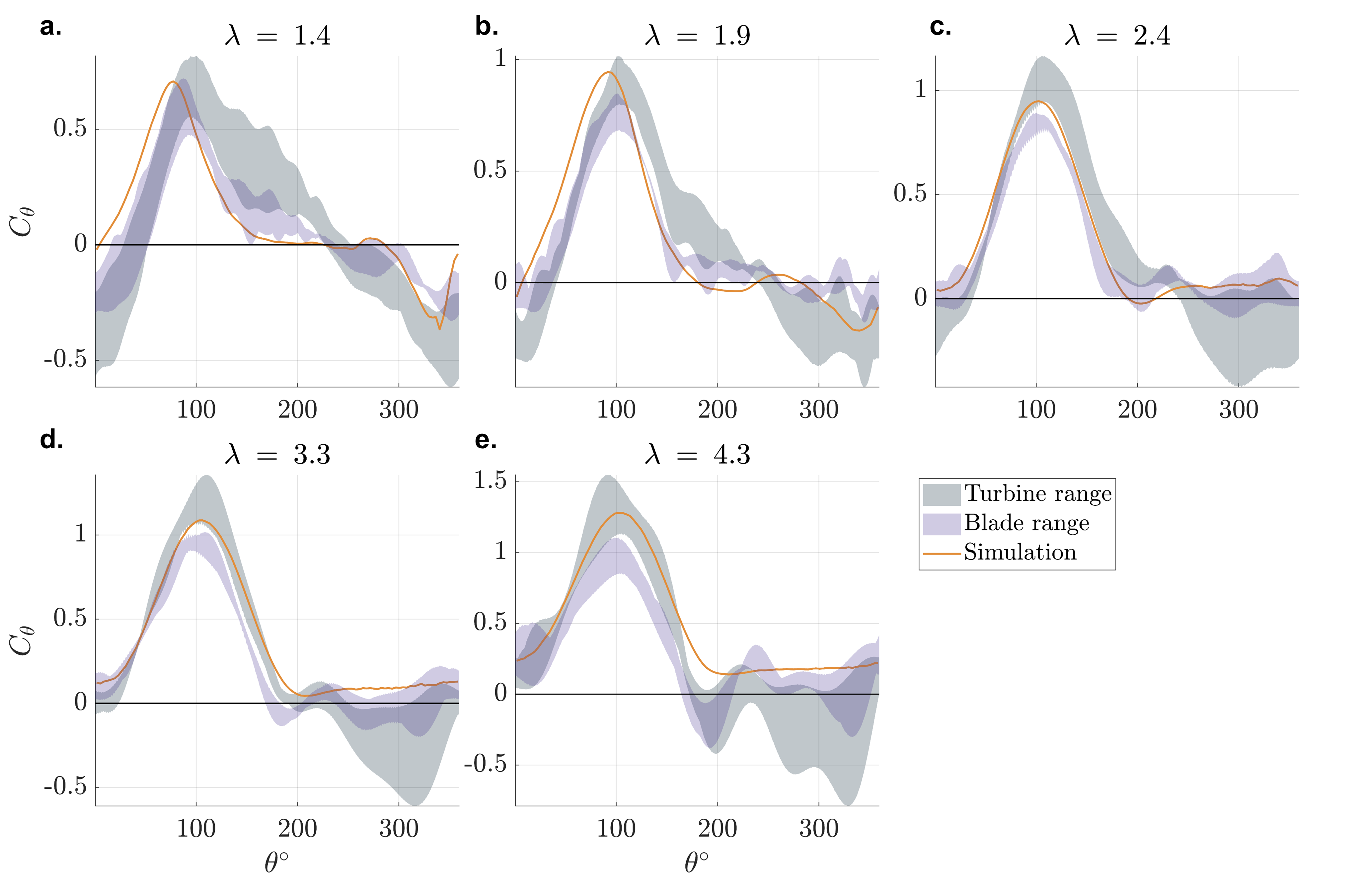}
    \caption{Phase-averaged tangential force coefficients. The colored area corresponds to the phase-averaged spread between the different shaft configurations (Figure \ref{shafts}).}
    \label{tantorquecomp}
\end{figure*}

\begin{figure*}[t!]
    \centering
    \includegraphics[width=1\linewidth]{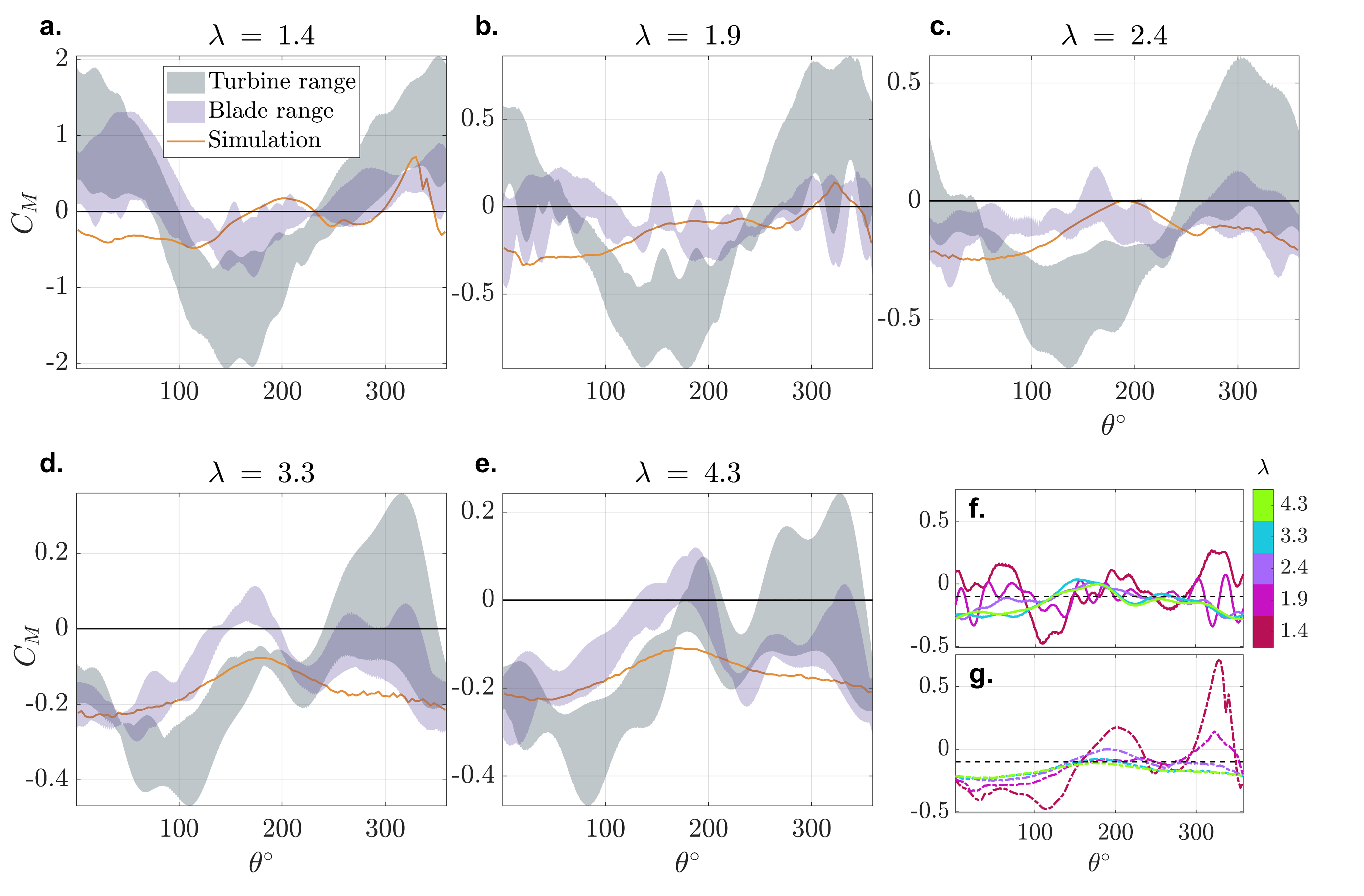}
    \caption{(a-e) Phase-averaged pitching moment coefficient. The colored area corresponds to the phase-averaged spread between the different shaft configurations (Figure \ref{shafts}). (f) Experimental pitching moment as a function of $\lambda$ for the ``Baseline'' shaft and (g) the same from simulation. The pitching moment for a NACA 6418 foil under low-to-moderate angles of attack is shown for reference in (f) and (g) (dashed line, $C_M\approx-0.1$) \cite{foilcoeffs}.}
    \label{pitchcomp}
\end{figure*}

Given the strong experimental-simulation agreement for time- and phase-averaged torque, thrust, and lateral force, we now consider the phase-averaged coefficients contributing to torque generation (Equation \ref{torque from coeffs}). The tangential force and pitching moment coefficients are presented in Figures \ref{tantorquecomp} and \ref{pitchcomp}, respectively. Differences between the experimental turbine- and blade-level tangential force and pitching moment coefficients are sizeable. Similarly, the spread in these quantities is reduced at the blade level. This demonstrates the importance of isolating blade-level forces when estimating these terms from single-bladed experimental data. 

\begin{figure*}[t!]
    \centering
    \includegraphics[width=1\linewidth]{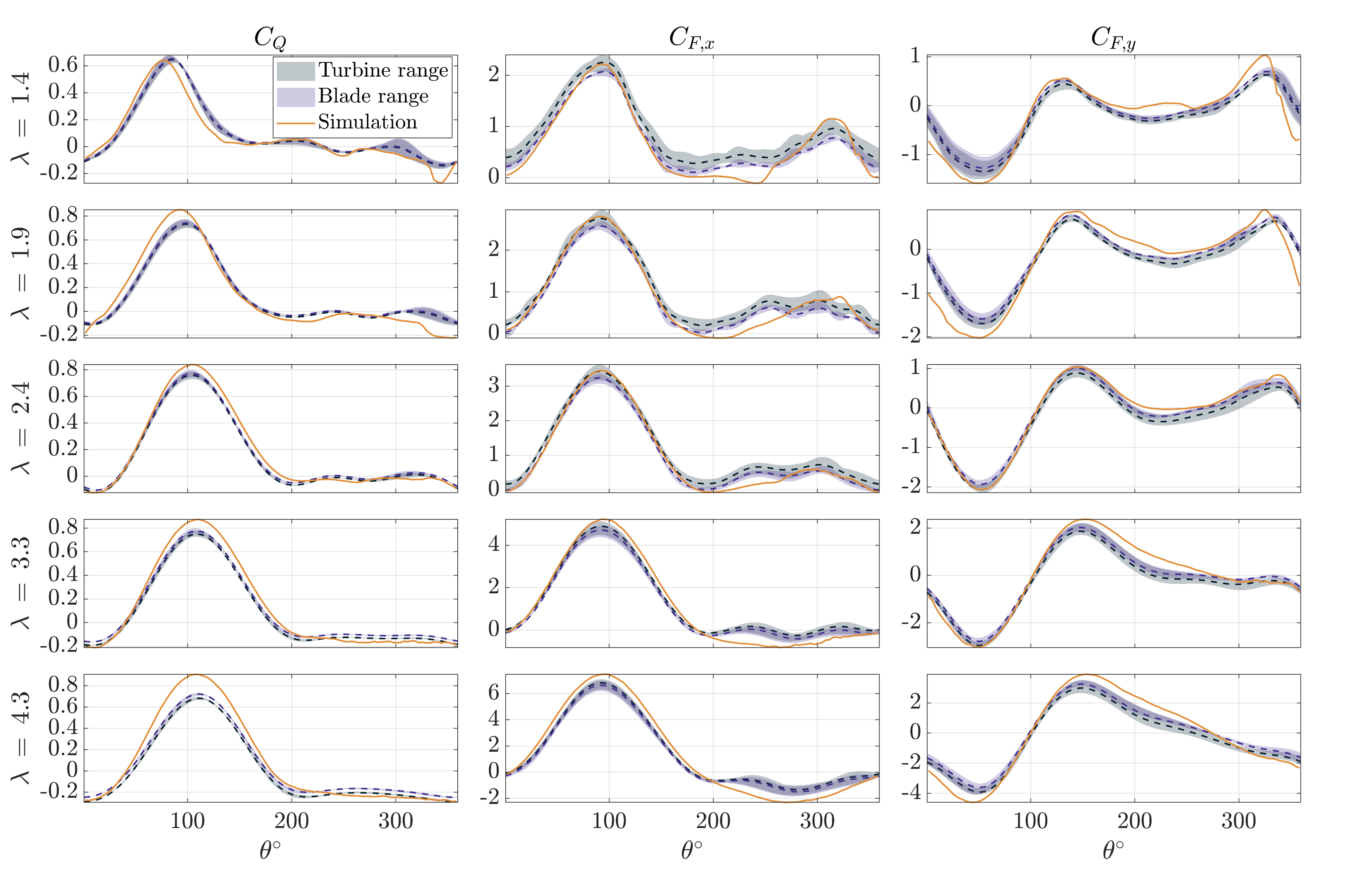}
    \caption{Phase-averaged, torque, thrust force, and lateral force coefficients. The colored area corresponds to the spread between phase-averages for the different shaft configurations and the dashed line represents the average of the different shaft configurations (Figure \ref{shafts}).}
    \label{phase averages}
\end{figure*}

The blade-level experimental and simulation tangential force coefficients have good quantitative agreement at all phases (Figure \ref{tantorquecomp}). For both the experiments and simulations, the tangential force coefficient follows the torque coefficient (Figure \ref{phase averages}), but with an amplitude offset that increases with the tip-speed ratio. This is a consequence of the pitching moment contribution (Equation \ref{torque from coeffs}). 

The phase-averaged, blade-level pitching moment coefficient has fair quantitative agreement between experiments and simulations with the experiments having larger fluctuations (Figure \ref{pitchcomp}). As for the tangential force coefficient, agreement is better for blade-level estimates. There are regions where the simulation values are always negative while some shaft configurations have positive blade-level $C_M$ values. This occurs in the first half of the upstream sweep for $\lambda\:=\:1.4$ (a) and $\lambda\:=\:1.9$ (b). 

Because of the relatively large range in the estimated pitching moment from experiments, trends across tip-speed ratios are not initially obvious. For clarity, phase-averaged, blade-level pitching moments for the ``Baseline'' shaft and the simulation data are presented in Figure \ref{pitchcomp}f,g, respectively. As the tip-speed ratio increases, the pitching moment range throughout the rotation decreases and approaches a constant, negative value. This agrees with Le Fouest et al. \cite{sebstalldilema} and is consistent with decreased dynamic stall severity at higher tip-speed ratios. The trend towards a negative average pitching moment is likely a consequence of virtual camber and convergence towards steady-state aerodynamics. As the tip-speed ratio increases, induction increases and azimuthal variations in the relative velocity and angle of attack are reduced, particularly in the downstream sweep \cite{SnortlandDownstream}. There, the relative velocity converges to the tangential velocity, and the angle of attack is determined by the preset pitch angle and virtual incidence. 
The experimentally-estimated pitching moment in the downstream sweep at the highest tip-speed ratios closely matches that for a NACA 6418 foil at low-to-moderate angles of attack, $-4 \geq \alpha \leq 12$  (Figure \ref{pitchcomp}f). That foil has a profile similar to virtual camber at the infinite tip-speed ratio limit for the turbine (zero inflow formulation from \cite{MiglioreWolfe}): maximum camber of $\approx6\%$ occurring between the quarter and half chord locations. 

Overall, while there are differences between simulation and experiments, the agreement in blade-level torque and forces is encouraging. Further, time-average, blade-level thrust nearly collapses across all shaft configurations. The phase-averaged collapse in blade-level forcing between shaft configurations is imperfect but significant relative to azimuthal force and torque variations. These results support the suitability of the physics-based strategy for isolating blade-level torque and force components from measurements at the axis of rotation for a single-bladed turbine.

\section{Discussion}
\label{discussion}
\begin{figure*}[t!]
    \centering
    \includegraphics[width=1\linewidth]{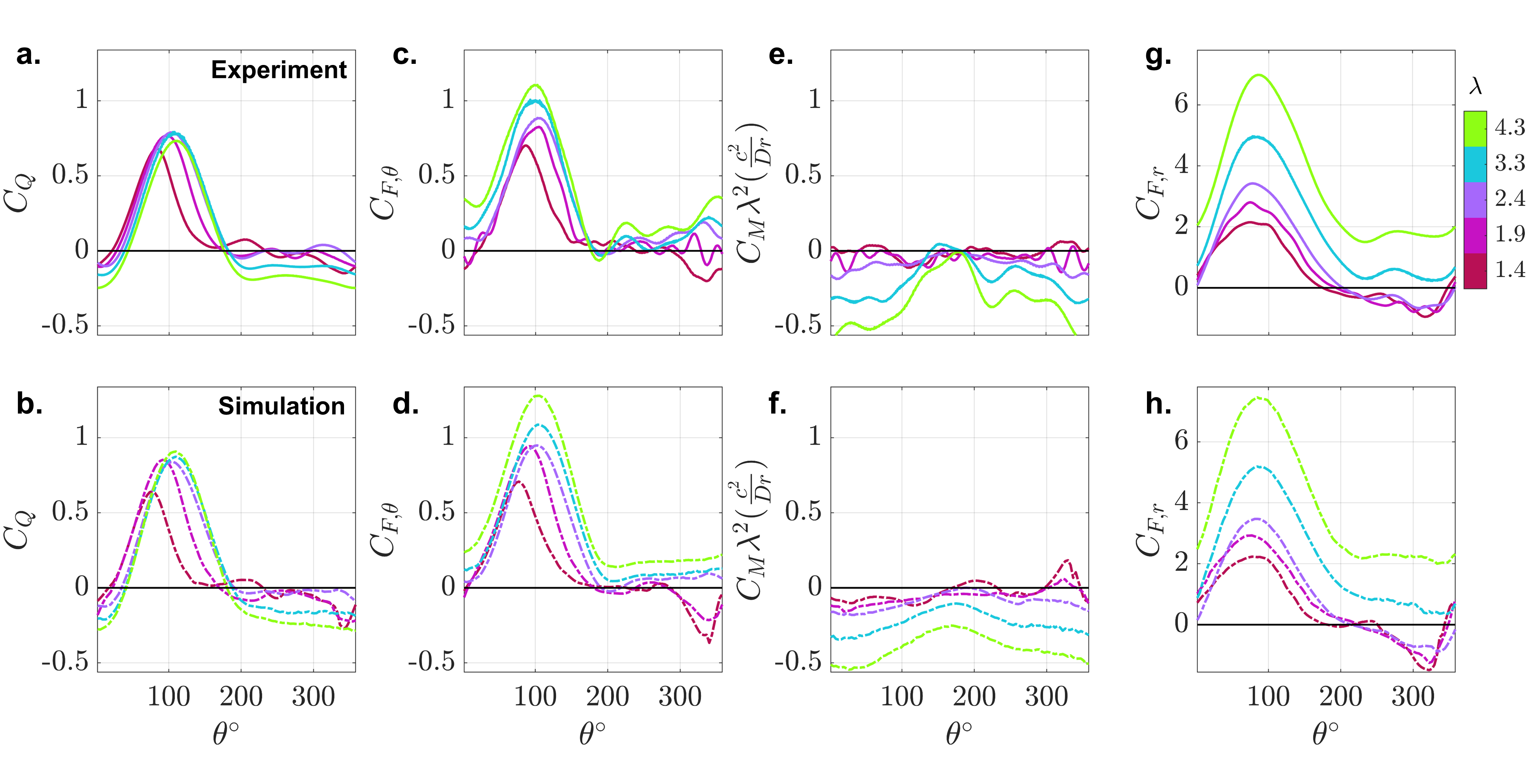}
    \caption{Blade-level (a,b) torque coefficient, (c,d) tangential force coefficient, (e,f) pitching moment contribution to torque, and (g,h) the normal force contribution for (top row)  experiments and (bottom row) simulation. Note: the $\lambda\:=\:1.3$ experiments are compared with simulation at $\lambda\:=\:1.4$.}
    \label{pitchcomp_standard}
\end{figure*}

Having established the general suitability of the superposition strategy, we now discuss overall trends, contributions, and implications of the estimated blade-level forces and pitching moment. We focus on the ``Baseline'' shaft configuration and the simulation phase-averaged results, as shown in Figure \ref{pitchcomp_standard}. Experimentally, the torque coefficient (a) shows little change with increasing tip-speed ratio during the upstream sweep, but becomes increasingly negative during the downstream sweep. In contrast, the tangential force coefficient (c) increases with the tip-speed ratio throughout the rotation and is positive in the downstream sweep for $\lambda\:>\:1.9$. The pitching moment contribution to torque (e) has a stronger dependence on $\lambda$ and is predominantly negative. Due to the $\lambda^2$ amplification, this term becomes increasingly detrimental at the higher tip-speed ratios even as $C_M$ converges towards a small, constant value. The normal force coefficient (g) increases continuously with the tip-speed ratio, likely because lift becomes more normal to the direction of rotation and increases in magnitude. The simulation results (b,d,f,h) show good agreement with these experimental trends, with the normal force coefficient having the best agreement. The maximum torque and tangential force coefficients in the upstream sweep are higher in simulation, while the experiments have more variable pitching moment contributions at the higher tip-speed ratios.

The normal force coefficient is positive throughout rotation for the highest two tip-speed ratios, signifying that, unlike for lower tip-speed ratios, the suction side never switches to the outside of the foil in the downstream sweep. Virtual camber and incidence likely explain this. At high tip-speed ratios, the angle of attack in the downstream sweep is largely determined by the preset pitch angle and virtual incidence. The virtual incidence for this geometry is $\approx 7^\circ$, resulting in an approximate $1^\circ$ angle of attack once the $-6^\circ$ preset pitch is accounted for. If we again consider NACA 6418 data, lift would point towards the center of rotation for angle of attack values $\geq-6^\circ$ \cite{foilcoeffs}. 
This may not be the case for all geometries, such as lower $c/r$ and more negative $\alpha_p$.

\subsection{Influence of the Pitching Moment on Turbine Performance}

\begin{figure*}[t!]
    \centering
    \includegraphics[width=1\linewidth]{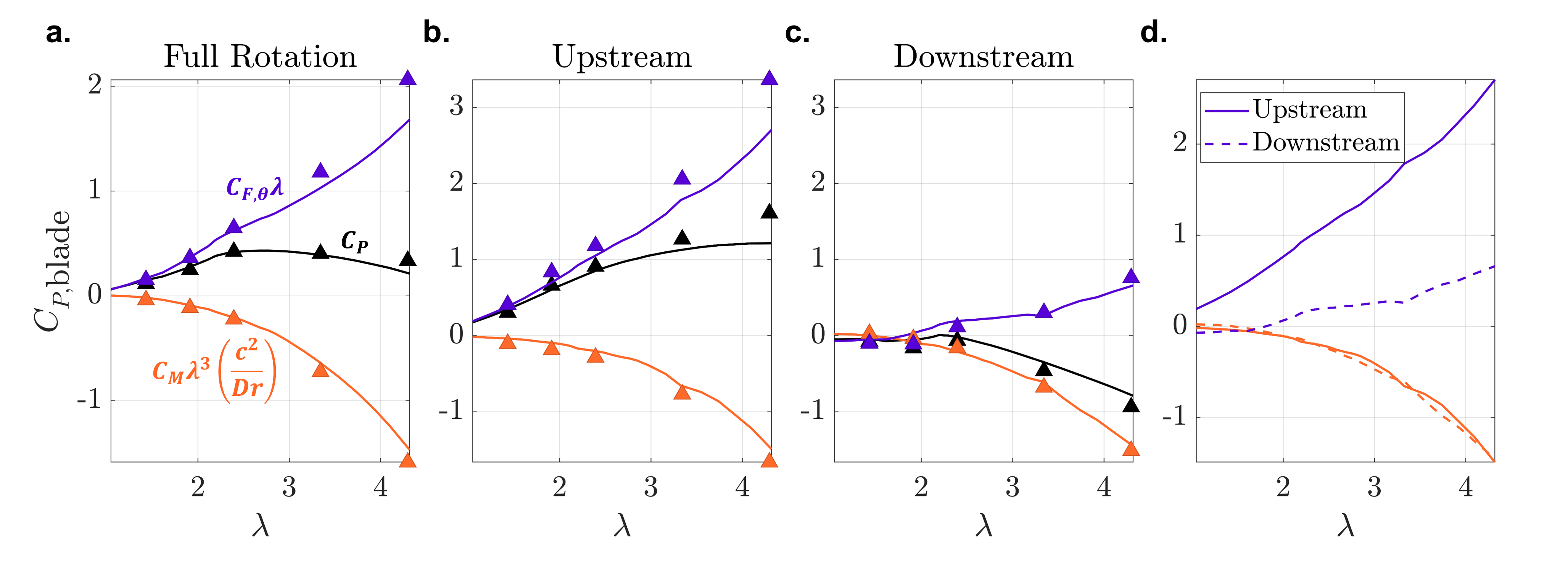}
    \caption{(a) Full rotation (time average), and (b) upstream and (c) downstream segment-averaged, blade-level performance compared to the corresponding tangential force and pitching moment contributions. The simulation results are plotted in (a-c) as the triangle markers. (d) Comparison between the tangential force and pitching moment performance contributions between the upstream and downstream sweeps.}
    \label{up down}
\end{figure*}

Snortland et al. \cite{SnortlandDownstream} demonstrated that continual degradation of downstream sweep performance determines the maximum time-averaged performance and the optimal tip-speed ratio. 
Using our new methodology, we can isolate the performance contributions from the tangential force (combination of the tangential projections of lift and drag) and the pitching moment for the upstream and downstream sweeps (Figure \ref{up down}). 
The upstream contribution is computed by averaging over 0-180$^\circ$, with an analogous range for the downstream contribution. 
The upstream and downstream contributions are scaled by 1/2 such that their sum is equal to the time-average. For the full rotation, upstream sweep, and downstream sweep, the tangential force contribution is offset by an increasingly detrimental pitching moment contribution (Figure \ref{up down}a-c). In fact, at the highest tip-speed ratio, which is approaching the freewheel condition, the pitching moment nearly cancels out the tangential force contribution. This suggests that the freewheel condition for some cross-flow turbines may be governed by different fluid dynamics than axial-flow turbines, where the tangential force goes to zero due to the offsetting influence of lift and drag forces. The pitching moment causes the performance degradation in the downstream sweep. However, the pitching moment is relatively similar for the upstream and downstream sweeps, meaning that the performance difference between these regions is primarily attributable to the tangential force. 
Overall, this highlights the importance of the pitching moment on cross-flow turbine performance; a stark contrast to Strickland's \cite{StricklandDMST} statement that the pitching moment is inconsequential. Going forward, blade element momentum models for cross-flow turbines (e.g., DMST) should include this term or explicitly prescribe lift and drag forces at the azimuthally-varying center of pressure. 

\subsection{Implications for Turbine Design}
The turbine- and blade-level force components have implications for structural design. As an example, for a turbine mounted on the seabed, thrust and lateral force are transferred to superstructures (i.e., support structures external to the turbine) and produce overturning moments that must be resisted by a foundation. Our results underscore the importance of considering phase-averaged force maximums in the design of cross-flow systems, as the time-averages often under-predict maximum loads. As an example, time-averaged thrust is substantially higher than the lateral force, but the maximum phase-average thrust and lateral force are comparable (Figure \ref{phase averages}). Information about the normal and tangential forces and the pitching moment also provides insight for blade design. For instance, the normal force does not contribute to power but is the largest component of blade-level fluid forcing and drives blade bending stresses and shear at the mounting points. For this turbine, the normal force coefficient can be up to six times the tangential force coefficient. 

Single-bladed turbines are unlikely to be utilized in practice, but studying their blade-level forces can inform the design of multi-bladed turbines. Moreover, the methods in this work may be applied to multi-bladed turbines with two caveats. First, since forces are measured at the center of rotation, ${F}^\star_{x,\textrm{blade}}$, ${F}^\star_{y,\textrm{blade}}$, and $Q^\star_\textrm{blade}$ contributions from individual blades are unknown. Second,  because of this, the total tangential force from all the blades remains unknown, and the pitching moment cannot be determined. However, applying the physics-based strategy to multi-bladed turbines could illuminate how aggregate blade-level thrust and lateral forcing differ with geometry and experimental conditions. Such analysis would decouple blade-level force trends from those of the support structures, and is especially useful for experiments where these are substantial. However, the aggregate blade-level force represents the sum of phase-shifted, oscillatory, individual blade forces that may constructively or destructively add. Therefore, it is unlikely the aggregate forces are representative of the range of forces experienced by an individual blade throughout the rotation. The number of blades will undoubtedly affect the force ranges experienced by an individual blade, but studying this requires direct blade-level force measurements (e.g., \cite{sebstalldilema}). Finally, in one regard, the validity of the physics-based strategy may improve when applied to multi-bladed devices since induction would increase with blade count \cite{rezaeiha2018solidity}, meaning the thrust loading on the shaft within the blade-span should be increasingly negligible.  

\section{Conclusion}

Consideration of blade-level forces and torques is key to demystifying the fundamental operation of cross-flow turbines and improving their structural design. To identify blade-level fluid forces and torques from experimental measurements at the axis of rotation, we define a physics-based superposition strategy. The precision of this method is demonstrated through the collapse in blade-level force and torque coefficients between turbine experiments consisting of four different shaft configurations and is in strong agreement with equivalent blade-only simulations. Most notably, the collapse in time-averaged blade-level thrust across shaft configurations is within the experimental standard deviation, and the simulation results largely overlap with this spread. Overall, the physics-based strategy is an empirical approach that works well for our single-bladed turbine geometry and is consistent with the flow physics. Factors such as higher blockage and non-ideal inflows may reduce its efficacy. In those cases, forcing on the blade-shaft connections could differ substantially between when the blades are present and when they are not. An experimental capability to concurrently measure turbine-level and blade-level forces and torques would be useful for establishing the limits of this method.

Isolation of experimental blade-level torque and forces facilitates the study of the normal force, tangential force, and pitching moment. The normal force is the largest component of blade-level forcing, is often six times greater than the tangential force, increases continuously with the tip-speed ratio, and is crucial for turbine structural design. Turbine performance is a function of the tangential force and pitching moment. The tangential force contribution increases continuously with tip-speed ratio and is largely responsible for differences between the upstream and downstream sweeps. In contrast, the often ignored pitching moment contribution is detrimental throughout the rotation and drives the continuous performance degradation in the downstream sweep, especially at high tip-speed ratios. These results underscore the general importance of studying the pitching moment and that failing to consider this component results in an over-prediction of turbine performance by a large margin. Notably, at the highest tip-speed ratios, the oppositional torque from the pitching moment nearly cancels out the substantial contribution from the tangential force, leading to the freewheel condition. This finding is important for improving analytical models of cross-flow turbines, such as double multiple streamtube theory. By the same token, reductions to the detrimental pitching moment term through geometry or control could improve cross-flow turbine performance.

\begin{appendices}
\section{Measurement of the Product of Rotating Mass and Radius to the Center of Mass}
\label{inertia}

\begin{figure*}[h!]
    \centering
    \includegraphics[width=1\linewidth]{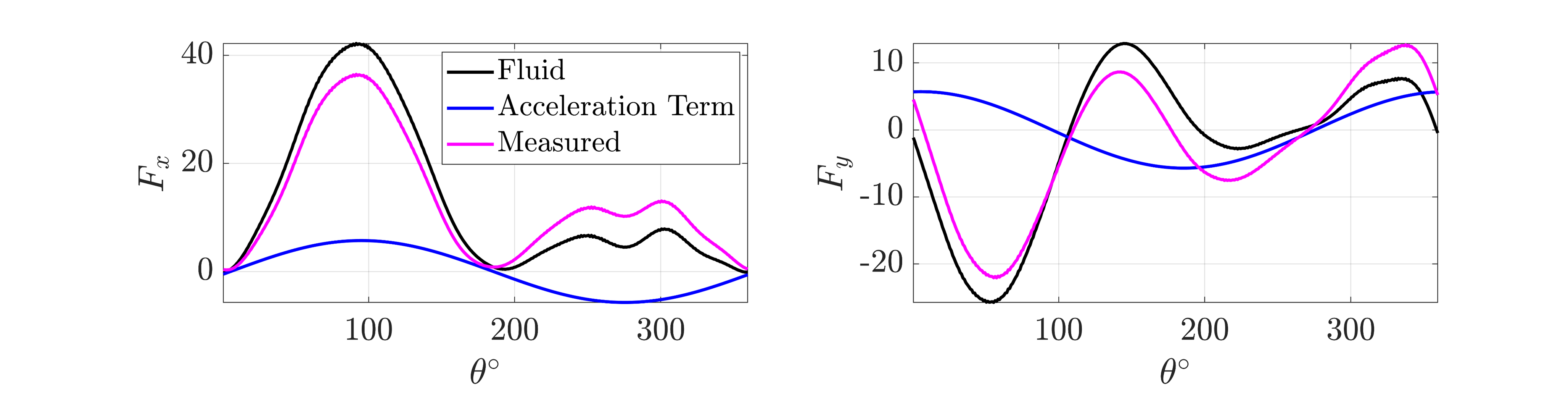}
    \caption{Phase-averaged fluid, acceleration, and measured force terms from Equations \ref{hydroX} and \ref{hydroY} for the ``Baseline'' full turbine configuration.}
    \label{force stack up}
\end{figure*}

To determine the product of the rotating mass and radius to the center of mass, $mr_g$, and the phase offset of the center of mass from the quarter-chord, $\delta\theta$, we run the turbine in quiescent air by commanding a range of sinusoidal $\omega$ profiles with different amplitudes and then apply Newton's second law. Assuming that the aerodynamic forces are negligible, the sum of forces in the streamwise, $x$, and cross-stream, $y$, directions are then 
\begin{equation}
\cancelto{0}{F^\star_{x}(t)}\:=\:mr_g[\omega(t)^2sin(\theta(t)-\delta\theta)-\dot{\omega}(t)cos(\theta(t)-\delta\theta)] + F_{\textrm{m},x}(t),
\end{equation}
and
\begin{equation}
\cancelto{0}{F^\star_{y}(t)}\:=\:mr_g[-\omega(t)^2cos(\theta(t)-\delta\theta)-\dot{\omega}(t)sin(\theta(t)-\delta\theta)] + F_{\textrm{m},y}(t).
\end{equation}

After smoothing the force data with a low-pass, zero-phase, Butterworth filter with a 10 Hz cutoff frequency, this system of two equations and two unknowns is solved using \emph{fsolve} in MATLAB. Even under constant speed control, the acceleration terms, $mr_g\omega^2\sin(\theta-\delta\theta)$ and $-mr_g\omega^2\cos(\theta-\delta\theta)$ in the $x$ and $y$ directions respectively (Equations \ref{hydroX} and \ref{hydroY}), are significant for a single-bladed turbine (Figure \ref{force stack up}).

\section{Superposition Strategy Details}
\label{superpos deets}

As discussed in Section~\ref{to blade}, the physics-based strategy involves measurements of the full turbine, support structures and shaft (i.e., blades removed), and shaft. For each of these experiments, the same laboratory flume static water level and pump rotation rate are used. Because the full turbine presents greater resistance to the flow, these identical settings result in slight variations in dynamic water depth and inflow velocity between the three configurations. This results in four subtleties.

First, the Froude number and Reynolds number vary slightly (< 1 \%) between test conditions. However, when the blades are removed, these changes are negligible in comparison to the reduction in blockage and the absence of induction. Second, force coefficients for each configuration should be calculated using the inflow for that specific test. This reduces error relative to superposition of phase-average dimensional quantities. Third, the discrete values of tip-speed ratios tested varies between experiments. We choose the tip-speed ratio for the full turbine as the reference condition and interpolate the measured tip-speed ratios for the support structures and shaft when calculating blade-level quantities through Equations \ref{blade level coeffs} and \ref{blade level coeffs torque}. Fourth, because of the changes in dynamic water depth, the submerged shaft length is slightly different between cases. The dynamic depth of the shaft-only case is represented as $H_\Upsilon$ and the dynamic depth for the turbine experiments is represented as $H_\Psi$. For thrust, the shaft force is scaled by the proportion of thrust on the shaft outside of the turbine span and the scaling factor $\boldsymbol{W_x}\:=\:\frac{H_\Psi-S'}{H_\Upsilon}$. For the lateral component, the shaft force is scaled only to account for the differences in $H$ for when the blades are present versus when they are absent and $\boldsymbol{W_y}\:=\:\frac{H_\Psi}{H_\Upsilon}$. This ratio is close to 1, so while it would be technically possible to adjust the pump rotation rate and static water depth in an \textit{ad hoc} manner to achieve identical inflow conditions between test configurations, this would be tedious and provide marginal benefit.

\begin{figure*}[t!]
    \centering
    \includegraphics[width=1\linewidth]{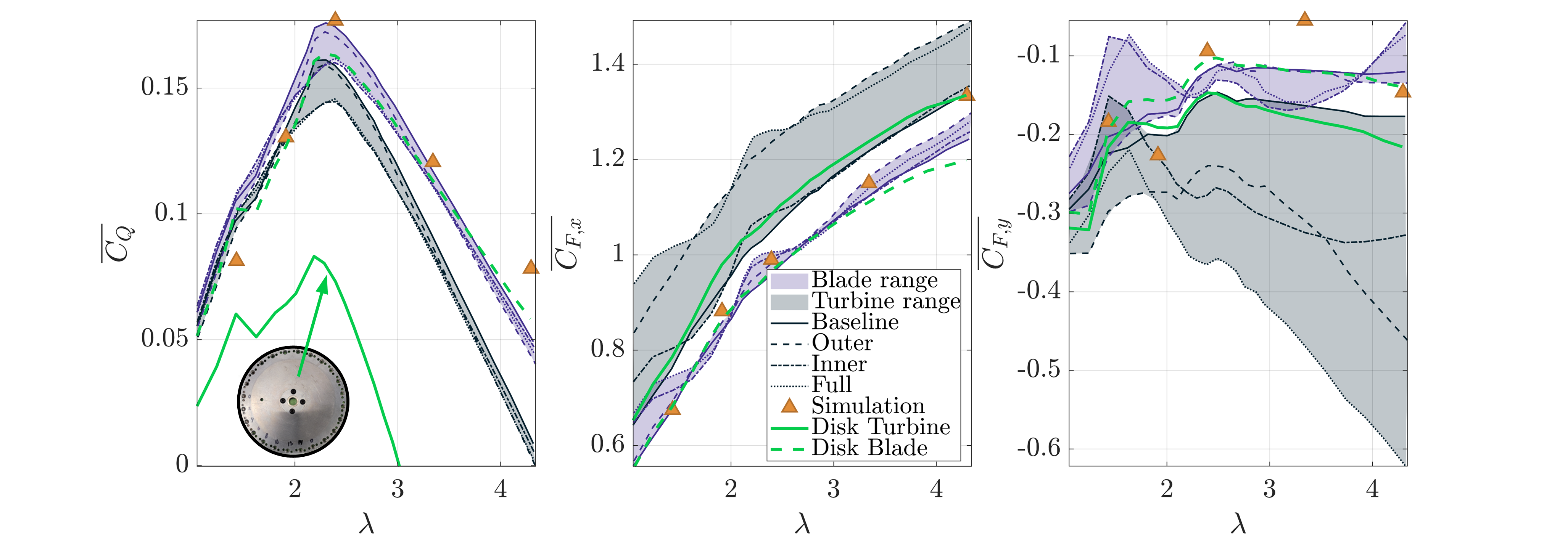}
    \caption{Time-averaged torque, thrust force, lateral force coefficients for the full turbine and the blades (from the physics-based strategy), for configurations using strut connections and a disk connection . The spread of the colored area corresponds to the range of the time-averaged values of the different shaft configurations (Figure \ref{shafts}).}
    \label{disk comparison}
\end{figure*}

Certain experimental campaigns may benefit from the use of non-ideal blade-shaft connections like end-plates. Such end plates have significantly higher parasitic torque losses than struts, but are cost-effective for sweeping through a broad set of geometric parameters, like preset pitch angle, blade count, and chord-to-radius ratio \citep{hunt2023parametric}. To investigate the efficacy of the physics-based strategy for a turbine with disk end-plates turbine, Figure \ref{disk comparison} compares blade-level force and torque computed from measurements of a turbine with 17.4 cm diameter disk end-plates to the present set of experiments and the blade-only simulation. Even though the turbine-level torque coefficient is much lower than the cases with struts, the resulting time-average blade-level torque and force coefficients are in close agreement with the strut results and simulation. This suggests that the physics-based strategy is robust to differences in support structures.

\begin{figure*}[t!]
    \centering
    \includegraphics[width=1\linewidth]{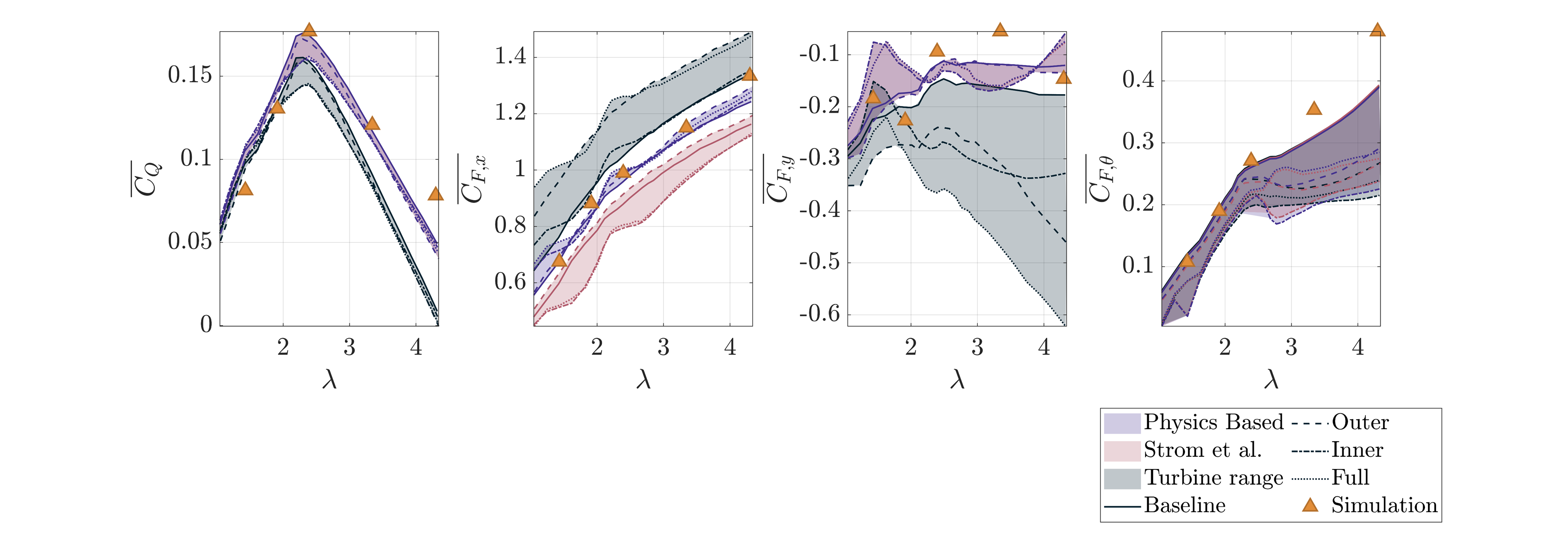}
    \caption{Time-averaged torque, thrust force, lateral force, and tangential force coefficients for the full turbine, the blades (from the physics-based and ``Strom et al.'' superposition strategies), and the simulations. The spread of the colored area corresponds to the range of the time-averaged values of the different shaft configurations (Figure \ref{shafts}).}
    \label{comparison}
\end{figure*}

Finally, in Figure \ref{comparison}, the physics-based approach is contrasted to the superposition strategy presented in Strom et al. \cite{stromsupports} where the blade-level coefficients are computed as 
\begin{equation}
\langle \boldsymbol{C}_{\textrm{blade}} \rangle\:=\:\langle \boldsymbol{C}_{\textrm{turb}} \rangle - \langle \boldsymbol{C}_{\textrm{sups}} \rangle.
\end{equation}
Here $\boldsymbol{C}$ represents an array of the thrust force, lateral force, and torque coefficients. Both superposition strategies decrease the magnitude of thrust and the lateral force relative to the turbine-level results. The torque calculation between the two strategies is identical, and the lateral force coefficient results are nearly identical. However, the physics-based strategy has substantially better agreement with the simulated time-averaged thrust coefficient and has a more physical basis. We also observe that the time-averaged tangential force coefficient is insensitive to the superposition strategy because of offsetting differences across phase.  

\end{appendices}

\section{Acknowledgments}
The authors thank the Alice C. Tyler Charitable Trust for supporting the research facility. The authors acknowledge the substantial contributions by Benjamin Strom, Hannah Ross, Erik Skeel, and Craig Hill to the development and upgrades of the experimental setup and code base and the insights provided by Carl Stringer at the onset of this work.
\subsection*{Funding}
Financial support was received from the United States Department of Defense Naval Facilities Engineering Systems Command and through the National Science Foundation Graduate Research Fellowship Program. 
\subsection*{Competing interests}
The authors have no relevant financial or non-financial interests to disclose.
\subsection*{Availability of data and materials}
Data and processing codes are available upon request.

\bibliography{References.bib}

\end{document}